\documentclass[aps,pra,twocolumn,showpacs,superscriptaddress]{revtex4-1}
\usepackage{amsmath}
\usepackage{amssymb}
\usepackage{graphicx}
\usepackage{braket}
\usepackage[utf8]{inputenc}
\usepackage[T1]{fontenc}

\makeatletter

\begin{document}

\title{Non-Markovian dynamics of mixed-state geometric phase of dissipative qubits}
\author{Wei Guo}
\affiliation{Zhejiang Institute of Modern Physics, Department of Physics,
Zhejiang University, Hangzhou 310027, China}
\author{Jian Ma}
\affiliation{Department of Chemistry, Massachusetts Institute of Technology,
77 Massachusetts Ave., Cambridge, Massachusetts 02139, USA}
\author{Xiaolei Yin}
\affiliation{Zhejiang
Institute of Modern Physics, Department of Physics, Zhejiang University,
Hangzhou 310027, China}
\author{Wei Zhong}
\affiliation{Zhejiang
Institute of Modern Physics, Department of Physics, Zhejiang University,
Hangzhou 310027, China}
\author{Xiaoguang Wang}
\email{xgwang@zimp.zju.edu.cn}
\affiliation{Zhejiang Institute of Modern
Physics, Department of Physics, Zhejiang University, Hangzhou 310027, China}
\begin{abstract}
We investigate the geometric phase of a two-level atom (qubit) coupled to a bosonic
reservoir with Lorentzian spectral density, and find that for the non-Markovian dynamics in which rotating-wave approximation (RWA) is performed,
geometric phase has a $\pi$-phase jump at the nodal point. However,
the exact result without RWA given by hierarchy equation of motion method shows that
there is no such a phase jump or nodal structure in geometric phase. Thus our results demonstrate that the counter-rotating terms significantly contribute to the geometric phase in multi-mode Hamiltonian under certain circumstances.
\end{abstract}
\pacs{03.65.Vf, 03.65.Yz, 42.50.Pq}
\maketitle

\section{Introduction}

Geometric phase, originally defined in a quantum system that undergoes an adiabatic,
unitary, and cyclic evolution~\cite{berry_quantal_1984}, is later realized to
be a holonomy effect in the Hilbert space~\cite{simon_holonomy_1983}. By
relaxing the superfluous assumptions, such as periodicity and adiabatic evolution, geometric phase was generalized to a much wider setting: for a cyclic but non-adiabatic evolution, Aharonov and
Anandan proved the existence of a Hamiltonian independent phase, which is called AA phase~\cite{anandan_geometry_1990}.
The classical counterpart, Pancharatnam phase, leads to
the generalization of geometric phase for almost arbitrary unitary
evolution~\cite{samuel_general_1988,wang_geometric_2013}. Geometric phase has been observed
in more than one
experiments~\cite{gp_experiment_optical,gp_experiment_nmr,gp_experiment_nmr2,gp_experiment_ec,gp_experiment_ec2}  , and it is also closely related to Bargmann invariants~\cite{rabei_bargmann_1999}.
However, when the initial state is orthogonal to the final state, the definition of
geometric phase breaks down, and Manini \textit{et al.} introduced a
complementary concept called off-diagonal geometric
phase~\cite{manini_off-diagonal_2000} to recover the phase information,
which was verified by Hasegawa \textit{et al.} in the neutron interference
experiment~\cite{hasegawa_off-diagonal_2001}. 

Another direction of generalization is to find the corresponding gauge invariant
phase of mixed states. Uhlmann~\cite{uhlmann_parallel_1986} proposed a quantum holonomy for a given path of density matrices under certain parallel transport condition. Sj\"{o}vist \textit{et al.} extended geometric phase to mixed states under unitary evolution~\cite{sjoqvist_geometric_2000} by considering feeding a mixed state into a Mach-Zehnder interferometer, which was verified by Ericsson \textit{et al.} by using single photon interferometry~\cite{ericsson_measurement_2005}. For mixed states undergoing
non-unitary evolution, Tong \textit{et al. }generalized the geometric phase
by applying the technique of purification~\cite{tong_kinematic_2004}. Motivated by~\cite{sjoqvist_geometric_2000}, Filipp and Sj\"{o}qvist~\cite{filipp_off-diagonal_2003} generalize off-diagonal geometric phase of mixed states under unitary evolution. Using Uhlmann's quantum holonomy,  Filipp and Sj\"{o}qvist~\cite{filipp_off-diagonal_2005} generalize off-diagonal geometric phase to non-unitary evolution. However there is some inconsistency between the two generalized off-diagonal geometric phase since the non-unitary one reduces to the unitary one only for density matrices with zero eigenvalues~\cite{filipp_off-diagonal_2005}, and the inconsistency~\cite{PhysRevLett.91.090405,slater_mixed_2002} also exists for mixed state geometric phase in Ref.~\cite{uhlmann_parallel_1986} and Ref.~\cite{sjoqvist_geometric_2000}. 

Geometric phase is a promising concept~\cite{wilczek_gp} in quantum
computation since it may provide a fault-tolerant way to perform
quantum operation~\cite{ekert_geometric_????}. One implementation of
quantum computation is through cavity QED, and Jaynes-Cummings (JC)
model~\cite{jaynes_comparison_1963} serves as a workhorse to cavity QED for
decades. JC model is successful in explaining various quantum optical
phenomena and meets with most experiments to a satisfactory
degree~\cite{Exploring_the_Quantum_2005}. However, the underlying
rotating-wave approximation(RWA) may produce incorrect predictions. An example
is the vacuum induced Berry phase~\cite{fuentes-guridi_vacuum_2002}, saying
that a qubit can acquire geometric phase even when there is no
photon in the cavity. Later it is proved to be merely a result of RWA and
disappears in Rabi model~\cite{larson_absence_2012}, where RWA is not performed. 
JC model and Rabi model all suppose that the electromagnetic field
inside the cavity is monochromatic (single mode), while in reality
the imperfection of cavity mirrors will broaden the spectral line, which serves
as a bosonic environment (bath), and there are already some works on how
geometric phase is affected by dephasing and dissipative
environment~\cite{yi_geometric_2005,yi_geometric_2006}.

Description of decoherence is a difficult problem, and only a few models can
be solved analytically. Hierarchy equation of motion method (HEOM), established by Tanimura
\emph{et
al.}~\cite{tanimura_1,*tanimura_2,*tanimura_3,*tanimura_4,*tanimura_5,*tanimura_6,*tanimura_7,*tanimura_8,*tanimura_9,tanimura_stochastic_2006},
is an exact numerical method that goes beyond Born and Markov approximation.
This method was first developed for system at finite temperature and being described by Drude spectrum. However, in quantum information process, the
qubits and devices are kept at very low temperature to protect entanglement. Ma
\emph{et al.} extended the HEOM to system
with Lorentz-type system-bath coupling spectrum at zero temperature~\cite{ma_entanglement_2012}, and it is found to be effective in computing
physical properties such as quantum Fisher
information~\cite{zhong_fisher_2013}.

In this work we reanalyze the behavior of geometric phase in
Ref.~\cite{chen_non-markovian_2010}, where a qubit is coupled to a cavity at
zero temperature with Lorentzian spectrum, and dipole interaction is assumed.
We show that their analysis is incomplete because singularities or nodal
structures~\cite{bhandari_singularities_2002} are left out, and we give an
upper bound to the existence of nodal structure. Moreover, by using
HEOM to obtain exact dynamics, we find nodal structures disappear and geometric phase is well defined in the whole parameter space.

The structure of this article is as follows. In Sec.~\ref{sec.gp}, we
introduce the geometric phase under non-unitary evolution, and explain how
Bargmann invariants is connected to the geometric phase. In
Sec.~\ref{sec.model}, we introduce the model of a qubit interacting with the
environment described by Lorentzian spectrum, and we derive the explicit expression of
geometric phase under RWA, where nodal structures are analyzed with special
attention. In Sec.~\ref{sec.hierarchy}, HEOM is
introduced and numerical results are shown and analyzed. A brief discussion
and summary are given in Sec.~\ref{sec.conclusion}.

\section{geometric phase and Bargmann invariants\label{sec.gp}}

In this section, we briefly review the definition of geometric phase for mixed states
under non-unitary evolution~\cite{tong_kinematic_2004}. When the evolution is unitary, cyclic and adiabatic, the geometric phase for a pure state is defined as
\begin{equation}\label{eq:berry phase}
\Phi_{g}=\arg\left[\braket{\psi\left(0\right)|\psi\left(T\right)} e^{-\int_{0}^{T}\braket{\psi\left(t\right)|\tfrac{\partial}{\partial t}|\psi\left(t\right)}\,\mathrm{d}t}\right],
\end{equation}
where $\braket{\psi\left(0\right)|\psi\left(T\right)}$ can be viewed as the total phase accumulated after $T$, with geometric phase part and dynamic phase part together, which is also called $\Phi_{\textrm{tot}}$. Multiply $\Phi_{\textrm{tot}}$ with~$\exp({-\int_{0}^{T}\braket{\psi\left(t\right)|\tfrac{\partial}{\partial t}|\psi\left(t\right)}\mathrm{d}t})$ offsets dynamic phase, leaves only the geometric phase. When the evolution is unitary only, Eq.~\eqref{eq:geo phase} reduces to the Berry phase under general settings~\cite{samuel_general_1988}.

The definition of geometric phase for mixed states under non-unitary evolution~\cite{tong_kinematic_2004} is
\begin{equation}\label{eq:geometric phase}
\Phi_{g}=\arg\bigg[\sum_i\sqrt{\varepsilon_{i}(0)\varepsilon_{i}(T)}\braket{i(0)|i(T)}e^{\int_{0}^{T}-\braket{i|\frac{\partial}{\partial t}|i}\mathrm{d}t}\bigg],
\end{equation}
where $\varepsilon_{i}(t)$~and~$\ket{i(t)}$ are $i$th eigenvalue and $i$th eigenvector of system's density matrix:~$\rho_{\mathrm{S}}\left(t\right)$, and $T$ is the total time of evolution. The meaning Eq.~\eqref {eq:geometric phase} can be understood as that the geometric phase of a state under non-unitary evolution is the weighted sum of geometric phase accumulated on each eigenstate of the initial state.
When evolution is unitary, Eq.~\eqref{eq:geometric phase} reduces to the
expression of geometric phase for mixed states under unitary
evolution~\cite{sjoqvist_geometric_2000}. If system is prepared at pure
state, then without loss of generality, we can assume that $\varepsilon_1(0)=1$
while the rest eigenvalues are all zero, and Eq.~\eqref{eq:geometric phase}
reduces to
\begin{equation}\label{eq:geo phase}
\Phi_{g}=\arg\left[\braket{\psi\left(0\right)|\psi\left(T\right)} e^{-\int_{0}^{T}\braket{\psi\left(t\right)|\tfrac{\partial}{\partial t}|\psi\left(t\right)}\,\mathrm{d}t}\right],
\end{equation}
which looks exactly the same as Eq.~\eqref{eq:berry phase} while the evolution here is non-unitary and non-cyclic. In Eq.~\eqref{eq:geo phase}, $\ket{\psi\left(t\right)}$ is the eigenvector of $\rho_\mathrm{S}\left(t\right)$, corresponding to eigenvalue $\varepsilon_1(t)$.
$\sqrt{\varepsilon_1\left(0\right)\varepsilon_1\left(T\right)}$ is omitted
because it is real and larger than $0$, thus has no effect under operation of taking out the phase. 

Using Taylor expansion, the integral on the exponential function in \eqref{eq:geo phase} can be
evaluated, to the first order, as
\begin{align}
e^{-\int_{0}^{T}\braket{\psi\left(t\right)|\tfrac{\partial}{\partial t}|\psi\left(t\right)}\mathrm{d}t}
 & \approx\prod_{i=1}^{N}e^{-\braket{\psi\left(t\right)|\frac{\partial}{\partial t}|\psi\left(t\right)}|_{t=t_i}\delta t}\nonumber\\
 & \approx\prod_{i=1}^{N}\left[1-\braket{\psi\left(t\right)|\frac{\partial}{\partial t}|\psi\left(t\right)}\big|_{t=t_i}\delta t\right]\nonumber\\
 & \approx\prod_{i=2}^{N}\braket{\psi\left(t_i\right)|\psi\left(t_{i-1}\right)},
\end{align}
where $t_1=0$,~$t_N=T$, ~$t_{i+1}=t_{i}+\delta t$ and $N=T/\delta t$. In the first line we approximate the integral using a summation of $N$ members, while the expressions in second line is the result of Taylor expansion and in third line we assume that $\ket{\psi(t)}$ is continuous, both are accurate to the first order of $\delta t$, thus the approximately equals are used. It is noted that in the last line, the range is changed from $[1,\,N]$ to $[2,\,N]$ due to the requirement of continuity. Then Eq.~\eqref{eq:geo phase} can be rewritten as a consecutive inner product:
\begin{eqnarray}
\Phi_g& \approx \arg\big[&\braket{\psi(T)|\psi(T-\delta t)}\ldots\braket{\psi(t_i)|\psi(t_{i-1})}\nonumber \\
                 &  & \ldots\braket{\psi(\delta t)|\psi(0)}\braket{\psi(0)|\psi(T)}\big],\label{eq:bargamann invariant}
\end{eqnarray}
the consecutive inner product under the operation $\arg$ is the complex conjugate of $n$-vertex Bargmann invariants~\cite{mukunda_bargmann_2001}. As $\delta t$ becomes infinitely small, the approximately equal in Eq.~\eqref{eq:bargamann invariant} can be replaced by equal, hence Bargmann invariannt and geometric phase are directly related in this limit~\cite{simon_bargmann_1993,phase_tracking}.

Compared with Eq.~\eqref{eq:geo phase}, Bargmann invariants expression is
 easy for numerical calculation. When finding the eigenvectors for a
matrix on computer, an arbitrary phase factor
$\chi\left(|\chi|=1\right)$ is attached, which leads to
anomalous behavior after the differential operation in Eq.~\eqref{eq:geo
phase}. Instead, there is no differentiation and integration but only basic inner
product operation within the Bargmann invariants expression
Eq.~\eqref{eq:bargamann invariant}, and the eigenvector at different time $t$
comes in pairs so the phase factors cancelled~$(\chi\chi^*=1)$.

\section{Dynamics and Geometric phase of a single qubit in lossy cavity\label{sec.model}}

We consider a two-level atom coupled to a bosonic bath (cavity) at zero temperature,
and the Hamiltonian~\cite{quantum_optics_1997} is
\begin{equation}\label{eq:hamiltonian}
H=H_\mathrm{S}+H_{\mathrm{B}}+H_{\mathrm{I}}
\end{equation}
where 
\begin{equation}
H_{\mathrm{S}}=\omega_{0}\sigma_{+}\sigma_{-}
\end{equation} 
is the Hamiltonian of the two-level atom (with $\hbar=1$);
\begin{equation}
H_{\mathrm{B}}=\sum_k\omega_{k}a_{k}^{\dagger}a_{k}
\end{equation}
is the Hamiltonian of the bath, and 
\begin{equation}\label{eq:spin-boson}
H_\mathrm{I}=\sigma_x\sum_k g_k\left(a^\dagger_k+a_k\right)
\end{equation}
describes the interaction between atom and bath, while $g_{k}$ represents
coupling strength between atom and $k$th mode of the bath, and $g_k$ is real.

In interaction picture, $H_{\mathrm{I}}$ becomes
\begin{eqnarray}
H_{\mathrm{I}}\left(t\right) & =  \sum_{k}g_{k}\bigg(&\sigma_{+}a_{k}^{\dagger}e^{\mathrm{i}\left(\omega_{0}+\omega_{k}\right)t}+\sigma_{+}a_{k}e^{\mathrm{i}\left(\omega_{0}-\omega_{k}\right)t}\label{eq:interaction_picture_interaction}\\
 &  & +\sigma_{-}a_{k}^{\dagger}e^{-\mathrm{i}\left(\omega_{0}-\omega_{k}\right)t}+\sigma_{-}a_{k}e^{-\mathrm{i}\left(\omega_{0}+\omega_{k}\right)t}\bigg)\nonumber
\end{eqnarray}
Apply rotation-wave approximation is to ignore the rapid oscillation terms, i.e., the terms with frequency ~$\omega_0+\omega_k$. And it is equivalent to write the Hamiltonian in Eq.~\eqref{eq:hamiltonian} as 
\begin{equation}
H=H_{\mathrm{S}}+H_{\mathrm{B}}+\sum_{k}g_{k}\left(\sigma_{+}a_{k}+\sigma_{-}a_{k}^{\dagger}\right),\label{eq:hamiltonian with RWA}
\end{equation}
which is exactly solvable by solving Schr\"{o}dinger equation of the
whole system. We suppose that the
initial state is
\begin{equation}\label{eq:product state}
\ket{\psi(0)}=\left(c_{0}(0)\ket{0}_{\textrm{S}}+c_{1}(0)\ket{1}_{\textrm{S}}\right)\ket{0_{k}}_{\textrm{E}},
\end{equation}
where $\ket{0}$ and $\ket{1}$ represent spin up and spin down state of qubit and $\ket{0_k}_\mathrm{E}$ represent vacuum state, thus bath and system are at product state initially. 

Under the evolution of hamiltonian with RWA, i.e., Eq.~\eqref{eq:hamiltonian with RWA}, the total state of bath and qubit at time $t$ takes the form
\begin{eqnarray}
\ket{\psi(t)}&=&\left(c_{0}(t)\ket{0}_{\textrm{S}}+c_{1}(t)\ket{1}_{\textrm{S}}\right)\ket{0_{k}}_{\textrm{E}}\nonumber\\
&&+\sum_{k}c_k(t)\ket{0}_{\textrm{S}}\ket{0\ldots1_k\ldots0}_{\textrm{E}}.\label{eq:intermediate_state}
\end{eqnarray} 
The evolution of initial state can be understood as follows: $\ket{0}_{\textrm{S}}\ket{0}_{\textrm{E}}$ doesn't evolve, while $\ket{1}_{\textrm{S}}\ket{0}_{\textrm{E}}$ evolves into two states: $\ket{1}_{\textrm{S}}\ket{0}_{\textrm{E}}$ and $\ket{0}_{\textrm{S}}\ket{0\ldots1_k\ldots0}_{\textrm{E}}$.  

It seems that the infinite number of modes $k$ leads to an unnormalizable state, while Eq.~\eqref{eq:intermediate_state} is actually normalized. 
Apply Schr\"{o}dinger equation to get the explicit expression of the amplitude's evolution in the interaction picture:
\begin{eqnarray}
c_{0}\left(t\right)&=&c_{0},\nonumber\\
\dot{c_{1}}\left(t\right)&=&-\mathrm{i}\sum_{k=0}^{\infty}e^{\mathrm{i}\left(\omega_{0}-\omega_{k}\right)t}c_{k}\left(t\right),\label{eq:coefficent_explicit_expression}\\
\dot{c_k}\left(t\right)&=&-\mathrm{i}g_{k}^{*}e^{-\mathrm{i}\left(\omega_{0}-\omega_k\right)t}c_{1}\left(t\right).\nonumber
\end{eqnarray}
With Eq.~\eqref{eq:coefficent_explicit_expression}, we have
\begin{align}
 & \frac{\mathrm{d}}{\mathrm{d}t}\left(\left|c_{0}\right|^2+\left|c_{1}\right|^{2}+\sum_{k=0}^{\infty}\left|c_{k}\right|^{2}\right)\label{eq:probability_conservation}\\
= & c_{1}\dot{c}_{1}^{*}+\sum_{k=0}^{\infty}c_{k}\dot{c}_{k}^{*}+c.c\nonumber\\
= & 0\nonumber
\end{align}
Since $|c_0(0)|^2+|c_1(0)|^2+\sum|c_k(0)|^2=1$, the state after evolution is also normalized.

In the calculation of correlation function
\begin{equation}
C(t-\tau)=\sum_kg_kg^{*}_{k}e^{\mathrm{i}(\omega_0-\omega_k)(t-\tau)},
\end{equation}
assuming the bath's degrees of freedom is large and modes are closely spaced, we can replace the summation with integral 
\begin{equation}
C\left(t-\tau\right)= \int_{0}^{+\infty}J\left(\omega\right)e^{\mathrm{i}(\omega_0-\omega)\left(t-\tau\right)}\mathrm{d}\omega,
\end{equation}
and $J(\omega)$ is called spectral density.
The spectral density we assume here is the Lorentzian shape~\cite{breuer_open_system}
\begin{equation}
J\left(\omega\right)=\dfrac{1}{\pi}\dfrac{W^{2}\lambda}{\left(\omega_{0}-\omega\right)^{2}+\lambda^{2}},\label{eq:lorentz spectrum}
\end{equation}
where $W$ reflects
the coupling strength between atom and bath and is proportional to $|g_k|$, and $\lambda$ is the spectral width
at the resonance frequency $\omega_{0}$.
The correlation function $C$ under Lorentzian spectral density can be calculated as 
\begin{eqnarray}
C\left(t-\tau\right)=W^{2}\exp\left[-\lambda\left(t-\tau\right)\right],
\end{eqnarray}
here we replace the lower limit of integral from $0$ to $-\infty$, which is known as Weisskopf-Wigner approximation~\cite{quantum_optics_1997}. The parameter $\lambda$ not only defines the spectral width but is also
connected to the bath correlation time with the relation
$\tau_{c}=\lambda^{-1}$. Comparing with the time scale of the system:
$\tau_0=2\pi/\omega_0$, when $\tau_c\gg\tau_0$, the dynamics is non-Markovian; when $\tau_c\ll\tau_0$, the dynamics is Markovian~\cite{chen_non-markovian_2010}.

The explicit expression of $c_1(t)$ is~\cite{breuer_open_system}
\begin{eqnarray}
c_{1}\left(t\right) & = & c_1(0)e^{-\mathrm{i}\omega_{0}t}e^{-\tfrac{\lambda t}{2}}\left[\cosh\left(\dfrac{\Omega t}{2}\right)+\dfrac{\lambda}{\Omega}\sinh\left(\dfrac{\Omega t}{2}\right)\right]\nonumber \\
 & = & c_1(0)e^{-\mathrm{i}\omega_{0}t}f\left(t\right),\label{eq:solution to c}
\end{eqnarray}
where
\begin{equation}\label{eq:f(t) in rho}
f\left(t\right)=\begin{cases}
e^{-\lambda t/2}\left[\cosh\left(\frac{\Omega t}{2}\right)+\frac{\lambda}{\Omega}\sinh\left(\frac{\Omega t}{2}\right)\right], & \lambda^{2}>4W^{2},\\
e^{-\lambda t/2}\left(1+\frac{\lambda t}{2}\right), & \lambda^{2}=4W^{2},\\
e^{-\lambda t/2}\left[\cos\left(\frac{\Omega't}{2}\right)+\frac{\lambda}{\Omega'}\sin\left(\frac{\Omega't}{2}\right)\right], & \lambda^{2}<4W^{2}.
\end{cases}
\end{equation}
where $\Omega=\sqrt{\lambda^{2}-4W^{2}}$ and $\Omega'=\sqrt{4W^{2}-\lambda^{2}}$.
When $\lambda^{2}\geq4W^{2}$,
$f\left(t\right)$ decreases from $1$ to $0$ monotonically; when
$\lambda^{2}<4W^{2}$, $f\left(t\right)$ decreases from $1$ to $0$ while oscillating,
acting like a under-damped oscillator. $f\left(t\right)$ is real
regardless of $\lambda$ and $W$. As $\lambda$ approaches zero, $f(t)$ becomes $\cos Wt$.

\subsection{Analytical result under RWA\label{gp.rwa}}

\begin{figure}[!tp]
  \includegraphics[width=1\linewidth]{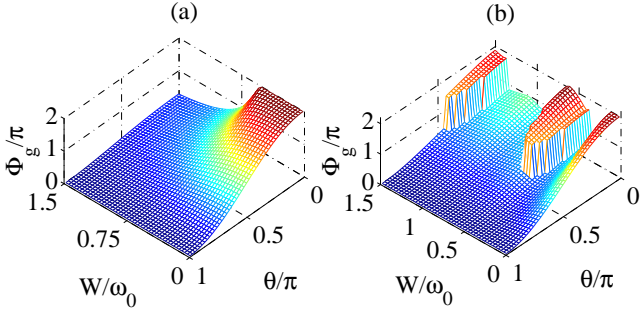}
   \caption{\label{fig:RWA}(Color online) Plots of $\Phi_g/\pi$ of Eq.~\eqref{eq:real GP} as a function of coupling strength $W$ and initial angle $\theta$.
 In (a), $\lambda=5\omega_0$, the dynamics is Markovian, and $\Phi_g$ decays monotonically with $W$ and $\theta$; in (b), $\lambda=0.05\omega_0$ thus the dynamics is in non-Markovian regime, and it's obvious that under RWA, $\Phi_g$ has a sudden change of $\pi$ at the vicinity of nodal point, where $\Phi_g$ is ill defined.}
\end{figure}

\begin{figure}[!tp]
  \centering\includegraphics[width=1\linewidth]{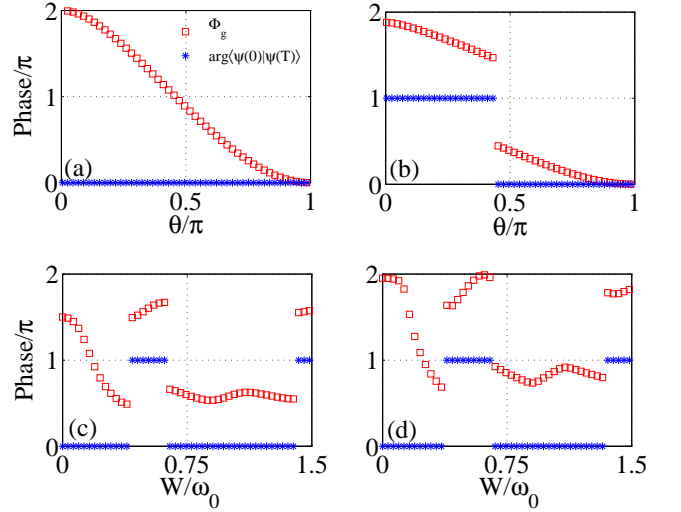}
    \caption{\label{fig:Intersection}(Color online) Plots of $\Phi_g/\pi$ of Eq.\eqref{eq:real GP} and $\textrm{arg}\braket{\psi(0)|\psi(T)}/\pi$ of Eq.~\eqref{eq:why} under RWA, with either initial angle $\theta$ or coupling strength $W$ fixed, $\lambda=0.05\omega_{0}$ is set for all sub-figures. Red square represents $\Phi_g$ and blue cross represents $\arg\braket{\Psi(0)|\Psi(T)}$.
In (a), coupling strength $W/\omega_0$ is fixed to $0.1$ and initial angle $\theta$ runs from $0$ to $\pi$; in (b) coupling strength is fixed to $W/\omega_0=0.5$ and $\theta$ varies. In (c), initial angle $\theta$ is kept as $\pi/3$ while coupling strength $W/\omega_0$ runs from $0$ to $1.5$, which is far beyond the strong coupling regime;
 In (d), initial angle $\theta$ is kept as $\pi/10$ and $W/\omega_0$ runs from 0 to 1.5. We observe that as long as $\textrm{arg}\braket{\psi(0)|\psi(T)}$ stays the same, $\Phi_g$ shows no ill behavior, as shown in (a), and vice versa, which is verified by the rest sub-figures.}
\end{figure}

We choose the initial state of qubit to be
\begin{equation}
\ket{\psi\left(0\right)}=\cos\dfrac{\theta}{2}\ket{1}+\sin\dfrac{\theta}{2}\ket{0},\label{eq:initial condition}
\end{equation}
and bath at vacuum state. After
tracing out the bath and employing Eq.~\eqref{eq:solution to c}, the reduced density matrix at time $t$ is
\begin{equation}\label{eq:reduced density matrix}
\rho_\mathrm{S}\left(t\right)=\left(\begin{array}{cc}
\cos^{2}\tfrac{\theta}{2}f^{2}\left(t\right) & \tfrac{\sin\theta}{2}f\left(t\right)e^{-\mathrm{i}\omega_{0}t}\\[15pt]
\tfrac{\sin\theta}{2}f\left(t\right)e^{\mathrm{i}\omega_{0}t} & 1-\cos^{2}\tfrac{\theta}{2}f^{2}\left(t\right)
\end{array}\right).
\end{equation}
The eigenvalues of $\rho_\mathrm{S}(t)$ are
\begin{equation}
\varepsilon_{\pm}\left(t\right)=\frac{1}{2}\pm\frac{1}{2}\sqrt{f^{2}\left(t\right)\sin^{2}\theta+\left(2f^{2}\left(t\right)\cos^{2}\dfrac{\theta}{2}-1\right)^{2}}, \label{eq:eigenvalue}
\end{equation}
with the corresponding eigenvectors:
\begin{equation}
\ket{\varepsilon_{\pm}\left(t\right)}=e^{-\mathrm{i}\omega_{0}t}\cos\Theta_{\pm}\ket{1}+\sin\Theta_{\pm}\ket{0},\label{eq:eigenvector +}
\end{equation}
where
\begin{subequations}
\begin{equation}
\label{eq:cosine theta}
\cos\Theta_{+}=\frac{\sin\theta f\left(t\right)}{\mathcal{N}_{+}(t)},\ 
\sin\Theta_{+}=\frac{2\left[\varepsilon_{+}-\cos^{2}\frac{\theta}{2}f^{2}\left(t\right)\right]}{\mathcal{N}_{+}(t)},
\end{equation}
and
\begin{equation}
\sin\Theta_{-}=\frac{\sin\theta f\left(t\right)}{\mathcal{N}_{-}(t)},\ 
\cos\Theta_{-}=\frac{2\left[\varepsilon_{-}-\cos^{2}\frac{\theta}{2}f^{2}\left(t\right)\right]}{\mathcal{N}_{-}(t)}.
\end{equation}
\end{subequations}
where 
\begin{equation}
\mathcal{N}_\pm(t)=\sqrt{4\left[\varepsilon_{\pm}-\cos^{2}\frac{\theta}{2}f^{2}\left(t\right)\right]^{2}+f^{2}\left(t\right)\sin^{2}\theta}
\end{equation} 
are normalizing factors.

\begin{figure}[!tp]
  \centering\includegraphics[width=1\linewidth]{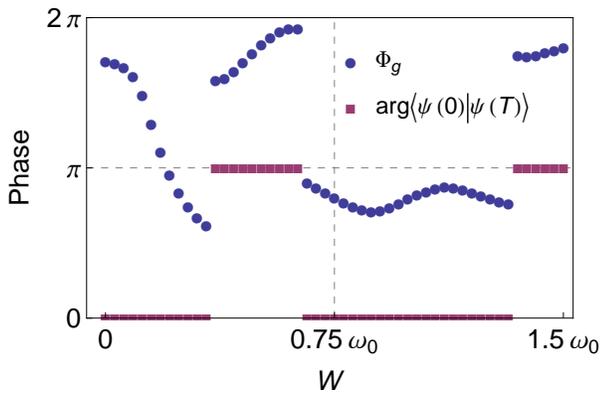}
    \caption{\label{fig:JC model gp}(Color online) Plots of $\Phi_g$ of Eq.\eqref{eq:real GP} and $\textrm{arg}\braket{\psi(0)|\psi(T)}$ under the evolution of Jaynes-Cummings model. Blue circles represents $\Phi_g$ and red squares represents $\arg\braket{\Psi(0)|\Psi(T)}$. Initial angle is fixed to $\pi/4$ with $W$ varying from zero to $1.5\omega_0$, and phase jump also exists.}
\end{figure}

Since system and bath are prepared in product state and system alone is in pure state initially, we apply Eq.~\eqref{eq:geo phase} to calculate the geometric phase acquired during~$T=2\pi/\omega_{0}$:
\begin{equation}\label{eq:gp of qubit under RWA}
\Phi_{g}=\textrm{arg}\left\{\cos\left[\dfrac{\theta}{2}-\Theta_{+}\left(T\right)\right]e^{\mathrm{i}\int_{0}^{T}\omega_{0}\cos^{2}\Theta\,\mathrm{d}t}\right\}.
\end{equation}
Equation~(\ref{eq:gp of qubit under RWA}) differs from the geometric phase
expression in Ref.~\cite{chen_non-markovian_2010} with a factor of
$\cos\left[\theta/2-\Theta_{+}\left(T\right)\right]$, which is simply $\arg\braket{\psi(0)|\psi(T)}$, and this factor is
crucial to the geometric phase and cannot be dropped. Actually, $\cos\left[\theta/2-\Theta_{+}\left(T\right)\right]$ may be negative or zero here, leading to a discontinuous $\Phi_g$. It's noted that as $\lambda$ approaches zero, substitute $f(t)=\cos{Wt}$ into Eq.~\eqref{eq:reduced density matrix}, then we recover the density matrix under evolution of JC model~\cite{jaynes_comparison_1963}, and phase jump still exists, as shown in Fig.~\ref{fig:JC model gp}.

A more explicit expression of geometric phase is
\begin{equation}
\Phi_{g}=\begin{cases}
\int_{0}^{T}\omega_{0}\cos^{2}\Theta_{+}\,\mathrm{d}t, & \cos\left[\tfrac{\theta}{2}-\Theta(T)\right]>0,\\
\textrm{undefined}, & \cos\left[\tfrac{\theta}{2}-\Theta(T)\right]=0,\\
\pi+\int_{0}^{T}\omega_{0}\cos^{2}\Theta_{+}\,\mathrm{d}t, & \cos\left[\tfrac{\theta}{2}-\Theta(T)\right]<0.
\end{cases}\label{eq:real GP}
\end{equation}
When $\cos\left[\theta/2-\Theta_{+}\right]$ changes sign, the
geometric phase has a sudden change of $\pi$; the point that
$\cos\left[\theta/2-\Theta_{+}\right]=0$ is called nodal point~\cite{filipp_off-diagonal_2003,bhandari_singularities_2002},
where the geometric phase has no definition. The discontinuity originates from the definition of geometric phase, e.g., $\arg(0.05)=0$ and $\arg(-0.05)=\pi$. In experiment, interference
visibility vanishes at nodal point.

Geometric phase with RWA as a function of initial angle $\theta$ and
coupling strength $W$ is plotted in Fig.~\ref{fig:RWA}, with Markovian case
($\tau_0/\tau_c=5$) in Fig.~\ref{fig:RWA}(a) and
non-Markovian case ($\tau_c/\tau_0=20$) in Fig.~\ref{fig:RWA}(b).
Coupling strength as large as $0.1\omega_0$ is enough to invalidate RWA,
therefore the region $W\in(0,~1.5\omega_0)$ we choose are adequate. When
dynamics is Markovian, $\Phi_g$ is continuous everywhere and decreases to 0
monotonically, which is believed to be caused by stronger dissipation and
shorter decoherence time~\cite{chen_non-markovian_2010}. When dynamics is
non-Markovian, $\Phi_g$ becomes discontinuous and has a phase jump of $\pi$, which is also known as nodal structure. We also observe that only if $\theta$ is smaller than $\pi/2$  will the phase jump occur, and this initial-state-dependent phenomenon will be explained at the end of this section.

The intersection of Fig.~\ref{fig:RWA} under non-Markovian dynamics is plotted
in Fig.~\ref{fig:Intersection}. It's clear that the phase jump of $\pi$ in
geometric phase is the direct result of sign change of
$\arg\braket{\psi\left(0\right)|\psi\left(T\right)}$. To restore the
phase information at nodal point requires the calculation of off-diagonal geometric
phase under non-unitary evolution~\cite{filipp_off-diagonal_2005}, however this is not easy due to the lack of explicit expression since the parallel transport condition in Ref.~\cite{filipp_off-diagonal_2005} is more abstract than the one in Ref.~\cite{tong_kinematic_2004}, and we will look into it in future studies.

Now we investigate how the difference between Markovian dynamics and non-Markovian dynamics causes the phase jump to happen. The explicit expression of
$\braket{\psi(0)|\psi(T)}$ is
\begin{equation}\label{eq:why}
\braket{\psi(0)|\psi(T)}=\frac{2\sin\frac{\theta}{2}\left\{\cos^{2}\frac{\theta}{2}\left[f\left(T\right)-f\left(T\right)^{2}\right]+\varepsilon_{+}\right\}}{\mathcal{N}_{+}(T)},
\end{equation}
since normalizing factor $\mathcal{N}_{+}(T)$ is always larger than 0, then the numerator is the only thing we need to concern. Since $\theta\in(0,\pi)$ then $\sin(\theta/2)>0$, and $\varepsilon_{+}>1/2$, so we focus on $f(T)-f(T)^2$. In Markovian dynamics regime, $\lambda\gg\omega_0$ and $c\left(t\right)$ decrease from 1 to 0 monotonically, thus $f(T)-f(T)^2$ is always larger than 0, then we have
\begin{equation}
\arg\braket{\psi(0)|\psi(T)}\equiv0
\end{equation} 
for Markovian dynamics regimes, and $\Phi_g$ is continuous. While in non-Markovian dynamics regimes, $\lambda^2-4W^2$ may be negative and $\Omega=\sqrt{\lambda^2-4W^2}$~becomes imaginary, resulting in a
decreasing and oscillating $f(t)$, thus it is possible for Eq.~\eqref{eq:why} to be zero or negative, thus we can conclude that $f<0$ is a necessary condition for $\Phi_g$ to become discontinuous.

The observation that $\Phi_g$ in Fig.~\ref{fig:RWA}~(b) becomes continuous after initial angle $\theta$ is sufficiently large ($\pi/2$) leads us to suspect whether there is a bound  on initial angle $\theta$ for geometric phase to be continuous under different bath setups, i.e, with different values of $\lambda$ and $W$. We know from the argument in previous paragraph that a phase jump must happen under non-Markovian dynamics, and the main reason is the oscillation of $\braket{\psi(0)|\psi(T)}$ from a positive number to an negative one, as shown in Eq.~\eqref{eq:why}. Now we derive the upper bound of initial angle $\theta$ for $\Phi_g$ to become discontinuous, i.e., for $\braket{\psi(0)|\psi(T)}$ to become negative. Use Eq.~\eqref{eq:why} and ignore $\mathcal{N}_+(T)$ together with $\sin\theta/2$ since it's always larger than $0$ for $\theta\in(0,\pi)$, we have
\begin{eqnarray}
\cos^{2}\frac{\theta}{2}\left(-\left|f\right|-f^{2}\right)+\varepsilon_{+} & \leq & 0\nonumber \\
\cos^{2}\frac{\theta}{2}\left(\left|f\right|+f^{2}\right) & \geq & \frac{1}{2}\nonumber \\
2\cos^{2}\frac{\theta}{2} & > & \frac{1}{2}\nonumber\\
\Rightarrow\theta&<&\theta_{\mathrm{C}}=\frac{2\pi}{3}\label{eq:loose_bound},
\end{eqnarray}
where in the first line we substitute $-|f|$ for $f$ since $f<0$ is necessary for the nodal structure; in the second line we use the fact that $\varepsilon_{+}\geq1/2$ and in
the third line $|f\left(t\right)|<1$. Therefore we conclude that $\theta<\theta_{\mathrm{C}}=2\pi/3$ is a necessary condition for geometric phase to be discontinuous under the evolution of Hamiltonian in Eq.~\eqref{eq:hamiltonian with RWA}; when $\theta>2\pi/3$, geometric phase must be continuous, regardless of $\lambda$ and $W$. The bound we derive in Eq.~\eqref{eq:loose_bound} covers the situation displayed in Fig.~\ref{fig:RWA}~(b), where $\Phi_g$ is discontinuous for $\theta<\pi/2$ and becomes continuous afterwards. It is noted that when $\theta=0$, geometric phase is ill defined regardless of $W$ and $\lambda$.

\subsection{\label{sec.hierarchy}Hierarchy equation of motion method}

\begin{figure}[!tp]
  \centering\includegraphics[width=1\linewidth]{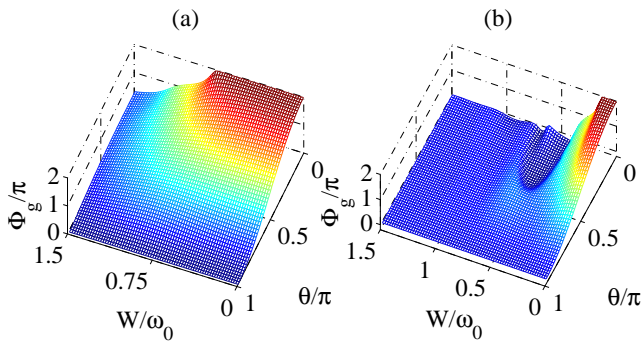}
    \caption{\label{fig:Hierarchy}(Color online) Numerical results, given by HEOM, of $\Phi_g/\pi$ without RWA as a function of coupling strength and initial angle.
     In (a), $\lambda=5\omega_0$, geometric phase under RWA decrease monotonically, just like the RWA case; in (b) $\lambda=0.05\omega_0$, geometric phase is continuous in whole parameter space, which is totally different from $\Phi_g$ under RWA.}
\end{figure}
\begin{figure}[htbp]
  \centering\includegraphics[width=0.8\linewidth]{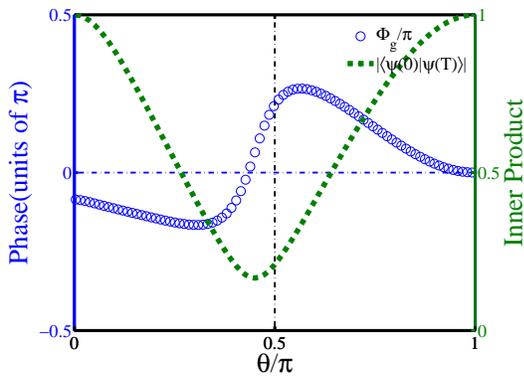}
    \caption{\label{fig:Hierarchy_intersection}(Color online) Numerical results of geometric phase (in the units of $\pi$) and the modulus of $\braket{\psi(0)|\psi(T)}$ without RWA versus initial angle,
    with $W=0.5\omega_0$ and $\lambda=0.05\omega_0$, i.e., within strong coupling regime and non-Markovian regime.
It is clear that the modulus never reaches $0$, thus $\Phi_g$ is continuous and well defined for all initial state.}
\end{figure}
\begin{figure}[htb]
  \centering\includegraphics[width=1\linewidth]{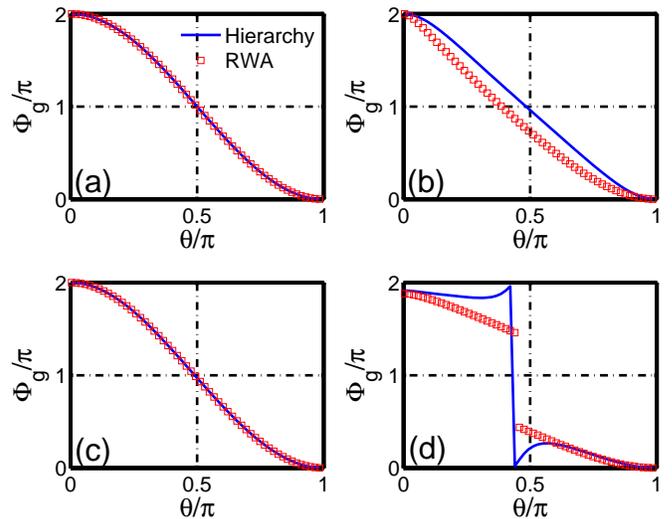}
    \caption{\label{fig:compare_Hier_RWA}(Color online) Comparison of geometric phase   with and without RWA as a function of initial angle $\theta$, both Markovian and non-Markovian regimes are considered. Blue lines represents geometric phase acquired through HEOM (without RWA) and red squares are geometric phase with RWA. We set $\lambda=5\omega_0$ for both (a) and (b), which lies in Markovian regime, and coupling strength $W=0.05\omega_0$ for (a) which is in the weak coupling regime, and $W=0.5\omega_0$ which is in the strong coupling regime for (b).  In (c) and (d), we set $\lambda=0.05\omega_0$ which lies in non-Markovian regime, and $W=0.05\omega_0$ which is in the weak coupling regime in (c), and $W=0.5\omega_0$ which is in the strong coupling regime in (d).}
\end{figure}
The sudden change of geometric phase is interesting and puzzling, but it maybe a misuse of RWA instead of physical reality. In Fig.~\ref{fig:Intersection}~(c) and \ref{fig:Intersection}~(d), $\Phi_g$ has a non-vanishing value even when $W\thicksim
 1\omega_0$, which is not likely since a large coupling strength indicates a quick decoherence and $\Phi_g$ should decrease to zero, implying that geometric phase under RWA may goes wrong. We also note that, in Fig.~\ref{fig:RWA}, the discontinuity lies in strong coupling strength area, which makes it suspicious. To answer these questions we need to know the exact dynamics of the system without RWA.

The density matrix of system, in the interaction picture, at time~$t$~is
\begin{equation}\label{eq:density matrix}
\rho_{\mathrm{tot}}\left(t\right)\equiv\mathcal{T}\exp\left[-i\int_{0}^{t}H_{\mathrm{SB}}^{\times}\left(\tau\right)\mathrm{d}\tau\right]\rho_{\mathrm{tot}}\left(0\right),
\end{equation}
where~$H_{\mathrm{SB}}\left(\tau\right)=e^{i\left(H_{\mathrm{S}}+H_{\mathrm{B}}\right)\tau}H_{\mathrm{SB}}e^{-i\left(H_{\mathrm{S}}+H_{\mathrm{B}}\right)\tau}$
is the interaction term in interaction picture. We also introduce the superscripts $^{\times}$ and
$^{\circ}$ to denote super-operators: $A^{\times}B\equiv[A,B]$ and
$A^{\circ}B\equiv\{A,B\}$, and $\mathcal{T}$ for time-ordering operator.

Hierarchy equation of motion method has two requirements: one is that system and bath are initially separable:
$\rho_\mathrm{tot}(0)=\rho_{\mathrm{S}}(0)\otimes\rho_\mathrm{B}(0)$, the
other is that system-bath interaction should be bilinear:
$H_{\mathrm{SB}}\left(\tau\right)=B\left(\tau\right)V\left(\tau\right)$,
where $B(\tau)$ and $V(\tau)$ are bath and system operators, respectively.
Both conditions are satisfied, as shown in Eqs.~\eqref{eq:hamiltonian} and~\eqref{eq:product state}. Trace out the bath, then we obtain
\begin{eqnarray}\label{eq:exact dynamics}
\rho_{\text{S}}^{\left(\text{I}\right)}\left(t\right) & = & \mathcal{T}\text{exp}\bigg\{-\int_{0}^{t}\mathrm{d}t_{2}\int_{0}^{t_{2}}\mathrm{d}t_{1}V\left(t_{2}\right)^{\times}[C^{\text{R}}\left(t_{2}-t_{1}\right)\nonumber \\
 & &\times V\left(t_{1}\right)^{\times} +\mathrm{i}C^{\text{I}}\left(t_{2}-t_{1}\right)V\left(t_{1}\right)^{\circ}\bigg\}\rho_{\text{S}}\left(0\right).
\end{eqnarray}
In Eq.~\eqref{eq:exact dynamics}, $V(0)=\sigma_{x}$ is the operator of qubit
in~$H_{\mathrm{SB}}$; $C^{\text{I}}$ and $C^{\text{R}}$ are imaginary and
real parts of the bath's time correlation function $\braket{B(t)B(0)}$ respectively, where
$B(0)=\sum_{k}g_{k}(a_{k}^{\dagger}+a_{k})$ is the operator of bath
in~$H_{\mathrm{SB}}$.

The principle of HEOM is to transform the operator
related integral Eq.~\eqref{eq:exact dynamics} to a set of ordinary
differential
equations~\cite{tanimura_stochastic_2006,ma_entanglement_2012,zhong_fisher_2013}:
\begin{eqnarray}\label{eq:hierarchy equation}
\frac{\partial}{\partial t}\varrho_{\vec{n}}\left(t\right) & = & -\left(\mathrm{i}H_{S}^{\times}+\vec{n}\cdot\vec{v}\right)\varrho_{\vec{n}}\left(t\right)-\mathrm{i}\sum_{k=1}^{2}V^{\times}\varrho_{\vec{n}+\vec{e_{k}}}\left(t\right)\nonumber \\
 & & -\mathrm{i}\frac{\lambda}{2}\sum_{k=1}^{2}n_{k}\left[V^{\times}+\left(-1\right)^{k}V^{\circ}\right]\varrho_{\vec{n}-\vec{e_{k}}}\left(t\right),
\end{eqnarray}
where $\vec{n}=(n_1,n_2)$, $\vec{e_1}=(1,0)$, $\vec{e_2}=(0,1)$ and
$\vec{v}=(\lambda-\mathrm{i}\omega_0,\lambda+\mathrm{i}\omega_0)$ are
auxiliary notations. The initial condition for Eq.~\eqref{eq:hierarchy equation}
is
\[
\varrho_{\vec{n}}\left(0\right)=\begin{cases}
\rho_{S}\left(0\right), & \textrm{for }n_{1}=n_{2}=0,\\
0, & \textrm{for }n_{1}>0,\ n_{2}>0.
\end{cases}
\]

The reduced density matrix $\rho\left(t\right)$ under the evolution of
Hamiltonian Eq.~\eqref{eq:hamiltonian}, is obtained by solving
Eq.~\eqref{eq:hierarchy equation} numerically. By calculating the evolution
of $\rho\left(t\right)$, whose initial state is
$\ket{\psi\left(0\right)}=\sin\tfrac{\theta}{2}\ket{1}+\cos\tfrac{\theta}{2}\ket{0}$,
we get an ordered series of eigenvectors
$\left\{\ket{\psi\left(t\right)}\right\}$, then substitute them into
Bargamann invariants Eq.~\eqref{eq:bargamann invariant}, and we obtain the
geometric phase without RWA.

Figure~\ref{fig:Hierarchy}~(a) displays $\Phi_g$ under Markovian dynamics without RWA, and it is similar to the one with RWA, as shown in Fig.~\ref{fig:RWA}, qualitatively. A more interesting result is that $\Phi_g$ within non-Markovian dynamics regimes without RWA is continuous everywhere, which is completely different. In Fig.~\ref{fig:Hierarchy}(b),  counter rotating terms are included, and nodal structures are gone. It is clear that under the evolution of full Hamiltonian and within non-Markovian dynamics regimes, geometric phase, as coupling strength becomes larger, decreases to $0$.

From Sec.~\ref{gp.rwa}, we know that with RWA, the discontinuity in $\Phi_g$ is the result of the sign change in $\braket{\psi(0)|\psi(T)}$, thus it's natural to investigate how $\braket{\psi(0)|\psi(T)}$ affects the $\Phi_g$ without RWA. The modulus of $\braket{\psi(0)|\psi(T)}$ and $\Phi_g$, obtained by HEOM, are shown in Fig.~\ref{fig:Hierarchy_intersection}, where $W=0.5\omega_0$ and $\lambda=0.05\omega_0$, i.e, within strong coupling and non-Markovian regime. We find that the modulus never reaches $0$, thus it cannot change sign under continuously changing parameters, which is $\theta$ here. $\Phi_g$ is continuous and tends to zero, as indicated in the first paragraph of Sec.~\ref{sec.hierarchy}. 

Rotating-wave approximation not only affects the geometric phase in non-Markovian regime, but it also affects geometric phase in Markovian regime, which is not evident at first sight. From Fig.~\ref{fig:compare_Hier_RWA}~(a)-(b), we find that when the coupling strength is weak, geometric phase is not affected by rotating-wave approximation regardless of Markovian or non-Markovian dynamic. Now we can observe easily how the counter rotating terms affects geometric phase. In Markovian regime, the deviation is not large, as shown in Fig.~\ref{fig:compare_Hier_RWA}~(c); however, in non-Markovian regime, the geometric phase with RWA deviates away from geometric phase without RWA, quantitatively and qualitatively. The sharp drop of geometric phase without RWA (blue line) in Fig.~\ref{fig:compare_Hier_RWA}~(d) is simply because we plot $\Phi_g\;\mathrm{mod}\;2\pi$ instead of $\Phi_g$, and this discontinuity of $2\pi$ can be removed by simply adding $2\pi$ back. Thus there is no nodal structure in geometric phase without RWA, while a discontinuity of $\pi$ in geometric phase with RWA cannot be removed.

 The reason that the behavior of GP in non-Markovian regime is significantly affected by RWA may be because RWA is not justified in describing non-Markovian dynamics. Intravia \textit{et al.} proved that for a harmonic oscillator the counter rotating terms has a significant contribution to the dynamic of system even in the weak coupling limit~\cite{intravaia_comparison_2003}; and M\"{a}kel\"{a} \textit{et al.} showed that for spin-boson model, RWA reduces non-Markovianity dramatically~\cite{makela_effects_2013}. Thus our result that the Geometric phase of a spin changes dramatically when coupled to a bosonic bath with Lorentzian spectrum in non-Markovian dynamic regime due to RWA is in agreement with the previous work, and we suggest that geometric phase may serve as an indicator of how non-Markovianity is affected by RWA. 

It is noted that Hamiltonian in Eq.~\eqref{eq:spin-boson} is the spin-boson model Hamiltonian after a $U=\exp\left(-{i\sigma_{\text{y}}3\pi}/2\right)$ rotation, and we know that the spin-boson model has a quantum phase transition where the ground state become localized as coupling strength goes over a critical value for Ohmic-like spectrum~\cite{leggett_dynamics_1987}. In this paper, we didn't see the effect of quantum phase transition on geometric phase, and the reason is that the initial state we choose, i.e., Eq.~\eqref{eq:product state}, is very different from the ground state of spin-boson model~\cite{chin_generalized_2011}, where one is product state while the other is entangled state, and fidelity of two states is small. Thus though the quantum phase transition can induce a dramatic change in the ground state and the geometric phase associated to the ground state~\cite{zhu_geometric_2008}, it does not have a significant effect on the geometric phase of Eq.~\eqref{eq:initial condition} here.   

\section{\label{sec.conclusion}Conclusion}

In summary, we have investigated the mixed-state geometric phase of a qubit, which is coupled to a bosonic bath with Lorentzian spectrum, with and without RWA. Under RWA, we find that there is discontinuity (nodal structures) in geometric phase when coupling strength is strong and dynamics is in non-Markovian regime. This result was not reported in the previous work~\cite{chen_non-markovian_2010} since the term $\arg\cos\left[\theta/2-\Theta_{+}\left(T\right)\right]$ was missing. How initial condition affects the nodal structure under non-Markovian dynamics is analyzed specifically, and a bound for discontinuity to disappear is given. 

Furthermore, with hierarchy equation of motion method, we calculated the geometric phase without RWA numerically. We find that, in the Markovian dynamics regime, the numerical result is in consonance with analytical result under RWA, thus the counter-rotating wave terms under such circumstances are not important. However, within non-Markovian dynamics regime, the behavior of geometric phase without RWA is quite different from the geometric phase with RWA. After taking the counter-rotating terms into consideration, the discontinuity in geometric phase disappears. Our results demonstrate that the counter-rotating terms significantly contribute to the geometric phase in multi-mode Hamiltonian when coupling strength is strong and within non-Markovian dynamics regime.

\section{Acknowledgment}
This work is supported by the NFRPC through Grant No. 2012CB921602, and the NSFC with Grants No. 11025527 and No. 10935010.
\bibliography{ref}

\begin{thebibliography}{53}%
\makeatletter
\providecommand \@ifxundefined [1]{%
 \@ifx{#1\undefined}
}%
\providecommand \@ifnum [1]{%
 \ifnum #1\expandafter \@firstoftwo
 \else \expandafter \@secondoftwo
 \fi
}%
\providecommand \@ifx [1]{%
 \ifx #1\expandafter \@firstoftwo
 \else \expandafter \@secondoftwo
 \fi
}%
\providecommand \natexlab [1]{#1}%
\providecommand \enquote  [1]{``#1''}%
\providecommand \bibnamefont  [1]{#1}%
\providecommand \bibfnamefont [1]{#1}%
\providecommand \citenamefont [1]{#1}%
\providecommand \href@noop [0]{\@secondoftwo}%
\providecommand \href [0]{\begingroup \@sanitize@url \@href}%
\providecommand \@href[1]{\@@startlink{#1}\@@href}%
\providecommand \@@href[1]{\endgroup#1\@@endlink}%
\providecommand \@sanitize@url [0]{\catcode `\\12\catcode `\$12\catcode
  `\&12\catcode `\#12\catcode `\^12\catcode `\_12\catcode `\%12\relax}%
\providecommand \@@startlink[1]{}%
\providecommand \@@endlink[0]{}%
\providecommand \url  [0]{\begingroup\@sanitize@url \@url }%
\providecommand \@url [1]{\endgroup\@href {#1}{\urlprefix }}%
\providecommand \urlprefix  [0]{URL }%
\providecommand \Eprint [0]{\href }%
\providecommand \doibase [0]{http://dx.doi.org/}%
\providecommand \selectlanguage [0]{\@gobble}%
\providecommand \bibinfo  [0]{\@secondoftwo}%
\providecommand \bibfield  [0]{\@secondoftwo}%
\providecommand \translation [1]{[#1]}%
\providecommand \BibitemOpen [0]{}%
\providecommand \bibitemStop [0]{}%
\providecommand \bibitemNoStop [0]{.\EOS\space}%
\providecommand \EOS [0]{\spacefactor3000\relax}%
\providecommand \BibitemShut  [1]{\csname bibitem#1\endcsname}%
\let\auto@bib@innerbib\@empty
\bibitem [{\citenamefont {Berry}(1984)}]{berry_quantal_1984}%
  \BibitemOpen
  \bibfield  {author} {\bibinfo {author} {\bibfnamefont {M.~V.}\ \bibnamefont
  {Berry}},\ }\href {\doibase 10.1098/rspa.1984.0023} {\bibfield  {journal}
  {\bibinfo  {journal} {Proc. R. Soc. Lond. A}\ }\textbf {\bibinfo {volume}
  {392}},\ \bibinfo {pages} {45} (\bibinfo {year} {1984})}\BibitemShut
  {NoStop}%
\bibitem [{\citenamefont {Simon}(1983)}]{simon_holonomy_1983}%
  \BibitemOpen
  \bibfield  {author} {\bibinfo {author} {\bibfnamefont {B.}~\bibnamefont
  {Simon}},\ }\href {\doibase 10.1103/PhysRevLett.51.2167} {\bibfield
  {journal} {\bibinfo  {journal} {Phys. Rev. Lett.}\ }\textbf {\bibinfo
  {volume} {51}},\ \bibinfo {pages} {2167} (\bibinfo {year}
  {1983})}\BibitemShut {NoStop}%
\bibitem [{\citenamefont {Anandan}\ and\ \citenamefont
  {Aharonov}(1990)}]{anandan_geometry_1990}%
  \BibitemOpen
  \bibfield  {author} {\bibinfo {author} {\bibfnamefont {J.}~\bibnamefont
  {Anandan}}\ and\ \bibinfo {author} {\bibfnamefont {Y.}~\bibnamefont
  {Aharonov}},\ }\href {\doibase 10.1103/PhysRevLett.65.1697} {\bibfield
  {journal} {\bibinfo  {journal} {Phys. Rev. Lett.}\ }\textbf {\bibinfo
  {volume} {65}},\ \bibinfo {pages} {1697} (\bibinfo {year}
  {1990})}\BibitemShut {NoStop}%
\bibitem [{\citenamefont {Samuel}\ and\ \citenamefont
  {Bhandari}(1988)}]{samuel_general_1988}%
  \BibitemOpen
  \bibfield  {author} {\bibinfo {author} {\bibfnamefont {J.}~\bibnamefont
  {Samuel}}\ and\ \bibinfo {author} {\bibfnamefont {R.}~\bibnamefont
  {Bhandari}},\ }\href {\doibase 10.1103/PhysRevLett.60.2339} {\bibfield
  {journal} {\bibinfo  {journal} {Phys. Rev. Lett.}\ }\textbf {\bibinfo
  {volume} {60}},\ \bibinfo {pages} {2339} (\bibinfo {year}
  {1988})}\BibitemShut {NoStop}%
\bibitem [{\citenamefont {Wang}\ and\ \citenamefont
  {Liu}(2013)}]{wang_geometric_2013}%
  \BibitemOpen
  \bibfield  {author} {\bibinfo {author} {\bibfnamefont {Z.~S.}\ \bibnamefont
  {Wang}}\ and\ \bibinfo {author} {\bibfnamefont {Q.}~\bibnamefont {Liu}},\
  }\href {\doibase 10.1016/j.physleta.2013.10.023} {\bibfield  {journal}
  {\bibinfo  {journal} {Physics Letters A}\ }\textbf {\bibinfo {volume}
  {377}},\ \bibinfo {pages} {3272} (\bibinfo {year} {2013})}\BibitemShut
  {NoStop}%
\bibitem [{\citenamefont {Tomita}\ and\ \citenamefont
  {Chiao}(1986)}]{gp_experiment_optical}%
  \BibitemOpen
  \bibfield  {author} {\bibinfo {author} {\bibfnamefont {A.}~\bibnamefont
  {Tomita}}\ and\ \bibinfo {author} {\bibfnamefont {R.~Y.}\ \bibnamefont
  {Chiao}},\ }\href {\doibase 10.1103/PhysRevLett.57.937} {\bibfield  {journal}
  {\bibinfo  {journal} {Phys. Rev. Lett.}\ }\textbf {\bibinfo {volume} {57}},\
  \bibinfo {pages} {937} (\bibinfo {year} {1986})}\BibitemShut {NoStop}%
\bibitem [{\citenamefont {Du}\ \emph {et~al.}(2007)\citenamefont {Du},
  \citenamefont {Zhu}, \citenamefont {Shi}, \citenamefont {Peng},\ and\
  \citenamefont {Suter}}]{gp_experiment_nmr}%
  \BibitemOpen
  \bibfield  {author} {\bibinfo {author} {\bibfnamefont {J.}~\bibnamefont
  {Du}}, \bibinfo {author} {\bibfnamefont {J.}~\bibnamefont {Zhu}}, \bibinfo
  {author} {\bibfnamefont {M.}~\bibnamefont {Shi}}, \bibinfo {author}
  {\bibfnamefont {X.}~\bibnamefont {Peng}}, \ and\ \bibinfo {author}
  {\bibfnamefont {D.}~\bibnamefont {Suter}},\ }\href {\doibase
  10.1103/PhysRevA.76.042121} {\bibfield  {journal} {\bibinfo  {journal} {Phys.
  Rev. A}\ }\textbf {\bibinfo {volume} {76}},\ \bibinfo {pages} {042121}
  (\bibinfo {year} {2007})}\BibitemShut {NoStop}%
\bibitem [{\citenamefont {Chen}\ \emph {et~al.}(2009)\citenamefont {Chen},
  \citenamefont {Hu}, \citenamefont {Chen},\ and\ \citenamefont
  {Du}}]{gp_experiment_nmr2}%
  \BibitemOpen
  \bibfield  {author} {\bibinfo {author} {\bibfnamefont {H.}~\bibnamefont
  {Chen}}, \bibinfo {author} {\bibfnamefont {M.}~\bibnamefont {Hu}}, \bibinfo
  {author} {\bibfnamefont {J.}~\bibnamefont {Chen}}, \ and\ \bibinfo {author}
  {\bibfnamefont {J.}~\bibnamefont {Du}},\ }\href {\doibase
  10.1103/PhysRevA.80.054101} {\bibfield  {journal} {\bibinfo  {journal} {Phys.
  Rev. A}\ }\textbf {\bibinfo {volume} {80}},\ \bibinfo {pages} {054101}
  (\bibinfo {year} {2009})}\BibitemShut {NoStop}%
\bibitem [{\citenamefont {Leek}\ \emph {et~al.}(2007)\citenamefont {Leek},
  \citenamefont {Fink}, \citenamefont {Blais}, \citenamefont {Bianchetti},
  \citenamefont {Göppl}, \citenamefont {Gambetta}, \citenamefont {Schuster},
  \citenamefont {Frunzio}, \citenamefont {Schoelkopf},\ and\ \citenamefont
  {Wallraff}}]{gp_experiment_ec}%
  \BibitemOpen
  \bibfield  {author} {\bibinfo {author} {\bibfnamefont {P.~J.}\ \bibnamefont
  {Leek}}, \bibinfo {author} {\bibfnamefont {J.~M.}\ \bibnamefont {Fink}},
  \bibinfo {author} {\bibfnamefont {A.}~\bibnamefont {Blais}}, \bibinfo
  {author} {\bibfnamefont {R.}~\bibnamefont {Bianchetti}}, \bibinfo {author}
  {\bibfnamefont {M.}~\bibnamefont {Göppl}}, \bibinfo {author} {\bibfnamefont
  {J.~M.}\ \bibnamefont {Gambetta}}, \bibinfo {author} {\bibfnamefont {D.~I.}\
  \bibnamefont {Schuster}}, \bibinfo {author} {\bibfnamefont {L.}~\bibnamefont
  {Frunzio}}, \bibinfo {author} {\bibfnamefont {R.~J.}\ \bibnamefont
  {Schoelkopf}}, \ and\ \bibinfo {author} {\bibfnamefont {A.}~\bibnamefont
  {Wallraff}},\ }\href {\doibase 10.1126/science.1149858} {\bibfield  {journal}
  {\bibinfo  {journal} {Science}\ }\textbf {\bibinfo {volume} {318}},\ \bibinfo
  {pages} {1889} (\bibinfo {year} {2007})},\ \bibinfo {note} {{PMID:}
  18033851}\BibitemShut {NoStop}%
\bibitem [{\citenamefont {M\"ott\"onen}\ \emph {et~al.}(2008)\citenamefont
  {M\"ott\"onen}, \citenamefont {Vartiainen},\ and\ \citenamefont
  {Pekola}}]{gp_experiment_ec2}%
  \BibitemOpen
  \bibfield  {author} {\bibinfo {author} {\bibfnamefont {M.}~\bibnamefont
  {M\"ott\"onen}}, \bibinfo {author} {\bibfnamefont {J.~J.}\ \bibnamefont
  {Vartiainen}}, \ and\ \bibinfo {author} {\bibfnamefont {J.~P.}\ \bibnamefont
  {Pekola}},\ }\href {\doibase 10.1103/PhysRevLett.100.177201} {\bibfield
  {journal} {\bibinfo  {journal} {Phys. Rev. Lett.}\ }\textbf {\bibinfo
  {volume} {100}},\ \bibinfo {pages} {177201} (\bibinfo {year}
  {2008})}\BibitemShut {NoStop}%
\bibitem [{\citenamefont {Rabei}\ \emph {et~al.}(1999)\citenamefont {Rabei},
  \citenamefont {Arvind}, \citenamefont {Mukunda},\ and\ \citenamefont
  {Simon}}]{rabei_bargmann_1999}%
  \BibitemOpen
  \bibfield  {author} {\bibinfo {author} {\bibfnamefont {E.~M.}\ \bibnamefont
  {Rabei}}, \bibinfo {author} {\bibnamefont {Arvind}}, \bibinfo {author}
  {\bibfnamefont {N.}~\bibnamefont {Mukunda}}, \ and\ \bibinfo {author}
  {\bibfnamefont {R.}~\bibnamefont {Simon}},\ }\href {\doibase
  10.1103/PhysRevA.60.3397} {\bibfield  {journal} {\bibinfo  {journal} {Phys.
  Rev. A}\ }\textbf {\bibinfo {volume} {60}},\ \bibinfo {pages} {3397}
  (\bibinfo {year} {1999})}\BibitemShut {NoStop}%
\bibitem [{\citenamefont {Manini}\ and\ \citenamefont
  {Pistolesi}(2000)}]{manini_off-diagonal_2000}%
  \BibitemOpen
  \bibfield  {author} {\bibinfo {author} {\bibfnamefont {N.}~\bibnamefont
  {Manini}}\ and\ \bibinfo {author} {\bibfnamefont {F.}~\bibnamefont
  {Pistolesi}},\ }\href {\doibase 10.1103/PhysRevLett.85.3067} {\bibfield
  {journal} {\bibinfo  {journal} {Phys. Rev. Lett.}\ }\textbf {\bibinfo
  {volume} {85}},\ \bibinfo {pages} {3067} (\bibinfo {year}
  {2000})}\BibitemShut {NoStop}%
\bibitem [{\citenamefont {Hasegawa}\ \emph {et~al.}(2001)\citenamefont
  {Hasegawa}, \citenamefont {Loidl}, \citenamefont {Baron}, \citenamefont
  {Badurek},\ and\ \citenamefont {Rauch}}]{hasegawa_off-diagonal_2001}%
  \BibitemOpen
  \bibfield  {author} {\bibinfo {author} {\bibfnamefont {Y.}~\bibnamefont
  {Hasegawa}}, \bibinfo {author} {\bibfnamefont {R.}~\bibnamefont {Loidl}},
  \bibinfo {author} {\bibfnamefont {M.}~\bibnamefont {Baron}}, \bibinfo
  {author} {\bibfnamefont {G.}~\bibnamefont {Badurek}}, \ and\ \bibinfo
  {author} {\bibfnamefont {H.}~\bibnamefont {Rauch}},\ }\href {\doibase
  10.1103/PhysRevLett.87.070401} {\bibfield  {journal} {\bibinfo  {journal}
  {Phys. Rev. Lett.}\ }\textbf {\bibinfo {volume} {87}},\ \bibinfo {pages}
  {070401} (\bibinfo {year} {2001})}\BibitemShut {NoStop}%
\bibitem [{\citenamefont {Uhlmann}(1986)}]{uhlmann_parallel_1986}%
  \BibitemOpen
  \bibfield  {author} {\bibinfo {author} {\bibfnamefont {A.}~\bibnamefont
  {Uhlmann}},\ }\href
  {http://www.sciencedirect.com/science/article/pii/0034487786900558}
  {\bibfield  {journal} {\bibinfo  {journal} {Reports on Mathematical Physics}\
  }\textbf {\bibinfo {volume} {24}},\ \bibinfo {pages} {229} (\bibinfo {year}
  {1986})}\BibitemShut {NoStop}%
\bibitem [{\citenamefont {Sj{\"o}qvist}\ \emph {et~al.}(2000)\citenamefont
  {Sj{\"o}qvist}, \citenamefont {Pati}, \citenamefont {Ekert}, \citenamefont
  {Anandan}, \citenamefont {Ericsson}, \citenamefont {Oi},\ and\ \citenamefont
  {Vedral}}]{sjoqvist_geometric_2000}%
  \BibitemOpen
  \bibfield  {author} {\bibinfo {author} {\bibfnamefont {E.}~\bibnamefont
  {Sj{\"o}qvist}}, \bibinfo {author} {\bibfnamefont {A.~K.}\ \bibnamefont
  {Pati}}, \bibinfo {author} {\bibfnamefont {A.}~\bibnamefont {Ekert}},
  \bibinfo {author} {\bibfnamefont {J.~S.}\ \bibnamefont {Anandan}}, \bibinfo
  {author} {\bibfnamefont {M.}~\bibnamefont {Ericsson}}, \bibinfo {author}
  {\bibfnamefont {D.~K.~L.}\ \bibnamefont {Oi}}, \ and\ \bibinfo {author}
  {\bibfnamefont {V.}~\bibnamefont {Vedral}},\ }\href {\doibase
  10.1103/PhysRevLett.85.2845} {\bibfield  {journal} {\bibinfo  {journal}
  {Phys. Rev. Lett.}\ }\textbf {\bibinfo {volume} {85}},\ \bibinfo {pages}
  {2845} (\bibinfo {year} {2000})}\BibitemShut {NoStop}%
\bibitem [{\citenamefont {Ericsson}\ \emph {et~al.}(2005)\citenamefont
  {Ericsson}, \citenamefont {Achilles}, \citenamefont {Barreiro}, \citenamefont
  {Branning}, \citenamefont {Peters},\ and\ \citenamefont
  {Kwiat}}]{ericsson_measurement_2005}%
  \BibitemOpen
  \bibfield  {author} {\bibinfo {author} {\bibfnamefont {M.}~\bibnamefont
  {Ericsson}}, \bibinfo {author} {\bibfnamefont {D.}~\bibnamefont {Achilles}},
  \bibinfo {author} {\bibfnamefont {J.~T.}\ \bibnamefont {Barreiro}}, \bibinfo
  {author} {\bibfnamefont {D.}~\bibnamefont {Branning}}, \bibinfo {author}
  {\bibfnamefont {N.~A.}\ \bibnamefont {Peters}}, \ and\ \bibinfo {author}
  {\bibfnamefont {P.~G.}\ \bibnamefont {Kwiat}},\ }\href {\doibase
  10.1103/PhysRevLett.94.050401} {\bibfield  {journal} {\bibinfo  {journal}
  {Phys. Rev. Lett.}\ }\textbf {\bibinfo {volume} {94}},\ \bibinfo {pages}
  {050401} (\bibinfo {year} {2005})}\BibitemShut {NoStop}%
\bibitem [{\citenamefont {Tong}\ \emph {et~al.}(2004)\citenamefont {Tong},
  \citenamefont {Sj{\"o}qvist}, \citenamefont {Kwek},\ and\ \citenamefont
  {Oh}}]{tong_kinematic_2004}%
  \BibitemOpen
  \bibfield  {author} {\bibinfo {author} {\bibfnamefont {D.~M.}\ \bibnamefont
  {Tong}}, \bibinfo {author} {\bibfnamefont {E.}~\bibnamefont {Sj{\"o}qvist}},
  \bibinfo {author} {\bibfnamefont {L.~C.}\ \bibnamefont {Kwek}}, \ and\
  \bibinfo {author} {\bibfnamefont {C.~H.}\ \bibnamefont {Oh}},\ }\href
  {\doibase 10.1103/PhysRevLett.93.080405} {\bibfield  {journal} {\bibinfo
  {journal} {Phys. Rev. Lett.}\ }\textbf {\bibinfo {volume} {93}},\ \bibinfo
  {pages} {080405} (\bibinfo {year} {2004})}\BibitemShut {NoStop}%
\bibitem [{\citenamefont {Filipp}\ and\ \citenamefont
  {Sj{\"o}qvist}(2003)}]{filipp_off-diagonal_2003}%
  \BibitemOpen
  \bibfield  {author} {\bibinfo {author} {\bibfnamefont {S.}~\bibnamefont
  {Filipp}}\ and\ \bibinfo {author} {\bibfnamefont {E.}~\bibnamefont
  {Sj{\"o}qvist}},\ }\href {\doibase 10.1103/PhysRevLett.90.050403} {\bibfield
  {journal} {\bibinfo  {journal} {Phys. Rev. Lett.}\ }\textbf {\bibinfo
  {volume} {90}},\ \bibinfo {pages} {050403} (\bibinfo {year}
  {2003})}\BibitemShut {NoStop}%
\bibitem [{\citenamefont {Filipp}\ and\ \citenamefont
  {Sj{\"o}qvist}(2005)}]{filipp_off-diagonal_2005}%
  \BibitemOpen
  \bibfield  {author} {\bibinfo {author} {\bibfnamefont {S.}~\bibnamefont
  {Filipp}}\ and\ \bibinfo {author} {\bibfnamefont {E.}~\bibnamefont
  {Sj{\"o}qvist}},\ }\href {\doibase 10.1016/j.physleta.2005.05.043} {\bibfield
   {journal} {\bibinfo  {journal} {Physics Letters A}\ }\textbf {\bibinfo
  {volume} {342}},\ \bibinfo {pages} {205} (\bibinfo {year}
  {2005})}\BibitemShut {NoStop}%
\bibitem [{\citenamefont {Ericsson}\ \emph {et~al.}(2003)\citenamefont
  {Ericsson}, \citenamefont {Pati}, \citenamefont {Sj\"{o}qvist}, \citenamefont
  {Br\"{a}nnlund},\ and\ \citenamefont {Oi}}]{PhysRevLett.91.090405}%
  \BibitemOpen
  \bibfield  {author} {\bibinfo {author} {\bibfnamefont {M.}~\bibnamefont
  {Ericsson}}, \bibinfo {author} {\bibfnamefont {A.~K.}\ \bibnamefont {Pati}},
  \bibinfo {author} {\bibfnamefont {E.}~\bibnamefont {Sj\"{o}qvist}}, \bibinfo
  {author} {\bibfnamefont {J.}~\bibnamefont {Br\"{a}nnlund}}, \ and\ \bibinfo
  {author} {\bibfnamefont {D.~K.~L.}\ \bibnamefont {Oi}},\ }\href {\doibase
  10.1103/PhysRevLett.91.090405} {\bibfield  {journal} {\bibinfo  {journal}
  {Phys. Rev. Lett.}\ }\textbf {\bibinfo {volume} {91}},\ \bibinfo {pages}
  {090405} (\bibinfo {year} {2003})}\BibitemShut {NoStop}%
\bibitem [{\citenamefont {Slater}(2002)}]{slater_mixed_2002}%
  \BibitemOpen
  \bibfield  {author} {\bibinfo {author} {\bibfnamefont {P.~B.}\ \bibnamefont
  {Slater}},\ }\href {\doibase 10.1023/A:1016199310947} {\bibfield  {journal}
  {\bibinfo  {journal} {Letters in Mathematical Physics}\ }\textbf {\bibinfo
  {volume} {60}},\ \bibinfo {pages} {123} (\bibinfo {year} {2002})}\BibitemShut
  {NoStop}%
\bibitem [{\citenamefont {Shapere}\ and\ \citenamefont
  {Wilczek}(1989)}]{wilczek_gp}%
  \BibitemOpen
  \bibfield  {author} {\bibinfo {author} {\bibfnamefont {A.}~\bibnamefont
  {Shapere}}\ and\ \bibinfo {author} {\bibfnamefont {F.}~\bibnamefont
  {Wilczek}},\ }\href@noop {} {\emph {\bibinfo {title} {Geometric Phases in
  Physics}}}\ (\bibinfo  {publisher} {World Scientific},\ \bibinfo {address}
  {Singapore},\ \bibinfo {year} {1989})\BibitemShut {NoStop}%
\bibitem [{\citenamefont {Ekert}\ \emph {et~al.}(2000)\citenamefont {Ekert},
  \citenamefont {Ericsson}, \citenamefont {Hayden}, \citenamefont {Inamori},
  \citenamefont {Jones}, \citenamefont {Oi},\ and\ \citenamefont
  {Vedral}}]{ekert_geometric_????}%
  \BibitemOpen
  \bibfield  {author} {\bibinfo {author} {\bibfnamefont {A.}~\bibnamefont
  {Ekert}}, \bibinfo {author} {\bibfnamefont {M.}~\bibnamefont {Ericsson}},
  \bibinfo {author} {\bibfnamefont {P.}~\bibnamefont {Hayden}}, \bibinfo
  {author} {\bibfnamefont {H.}~\bibnamefont {Inamori}}, \bibinfo {author}
  {\bibfnamefont {J.~A.}\ \bibnamefont {Jones}}, \bibinfo {author}
  {\bibfnamefont {K.~L.}\ \bibnamefont {Oi}}, \ and\ \bibinfo {author}
  {\bibfnamefont {V.}~\bibnamefont {Vedral}},\ }\href@noop {} {\bibfield
  {journal} {\bibinfo  {journal} {J. Mod. Opt}\ }\textbf {\bibinfo {volume}
  {47}},\ \bibinfo {pages} {2501} (\bibinfo {year} {2000})}\BibitemShut
  {NoStop}%
\bibitem [{\citenamefont {Jaynes}\ and\ \citenamefont
  {Cummings}(1963)}]{jaynes_comparison_1963}%
  \BibitemOpen
  \bibfield  {author} {\bibinfo {author} {\bibfnamefont {E.}~\bibnamefont
  {Jaynes}}\ and\ \bibinfo {author} {\bibfnamefont {F.}~\bibnamefont
  {Cummings}},\ }\href {\doibase 10.1109/PROC.1963.1664} {\bibfield  {journal}
  {\bibinfo  {journal} {P IEEE}\ }\textbf {\bibinfo {volume} {51}},\ \bibinfo
  {pages} {89 } (\bibinfo {year} {1963})}\BibitemShut {NoStop}%
\bibitem [{\citenamefont {Haroch}\ and\ \citenamefont
  {Raimond}(2005)}]{Exploring_the_Quantum_2005}%
  \BibitemOpen
  \bibfield  {author} {\bibinfo {author} {\bibfnamefont {S.}~\bibnamefont
  {Haroch}}\ and\ \bibinfo {author} {\bibfnamefont {J.-M.}\ \bibnamefont
  {Raimond}},\ }\href@noop {} {\emph {\bibinfo {title} {Exploring the
  Quantum}}}\ (\bibinfo  {publisher} {Oxford University Press},\ \bibinfo
  {address} {Oxford},\ \bibinfo {year} {2005})\BibitemShut {NoStop}%
\bibitem [{\citenamefont {Fuentes-Guridi}\ \emph {et~al.}(2002)\citenamefont
  {Fuentes-Guridi}, \citenamefont {Carollo}, \citenamefont {Bose},\ and\
  \citenamefont {Vedral}}]{fuentes-guridi_vacuum_2002}%
  \BibitemOpen
  \bibfield  {author} {\bibinfo {author} {\bibfnamefont {I.}~\bibnamefont
  {Fuentes-Guridi}}, \bibinfo {author} {\bibfnamefont {A.}~\bibnamefont
  {Carollo}}, \bibinfo {author} {\bibfnamefont {S.}~\bibnamefont {Bose}}, \
  and\ \bibinfo {author} {\bibfnamefont {V.}~\bibnamefont {Vedral}},\ }\href
  {\doibase 10.1103/PhysRevLett.89.220404} {\bibfield  {journal} {\bibinfo
  {journal} {Phys. Rev. Lett.}\ }\textbf {\bibinfo {volume} {89}},\ \bibinfo
  {pages} {220404} (\bibinfo {year} {2002})}\BibitemShut {NoStop}%
\bibitem [{\citenamefont {Larson}(2012)}]{larson_absence_2012}%
  \BibitemOpen
  \bibfield  {author} {\bibinfo {author} {\bibfnamefont {J.}~\bibnamefont
  {Larson}},\ }\href {\doibase 10.1103/PhysRevLett.108.033601} {\bibfield
  {journal} {\bibinfo  {journal} {Phys. Rev. Lett.}\ }\textbf {\bibinfo
  {volume} {108}},\ \bibinfo {pages} {033601} (\bibinfo {year}
  {2012})}\BibitemShut {NoStop}%
\bibitem [{\citenamefont {Yi}\ \emph {et~al.}(2005)\citenamefont {Yi},
  \citenamefont {Wang},\ and\ \citenamefont {Wang}}]{yi_geometric_2005}%
  \BibitemOpen
  \bibfield  {author} {\bibinfo {author} {\bibfnamefont {X.~X.}\ \bibnamefont
  {Yi}}, \bibinfo {author} {\bibfnamefont {L.~C.}\ \bibnamefont {Wang}}, \ and\
  \bibinfo {author} {\bibfnamefont {W.}~\bibnamefont {Wang}},\ }\href {\doibase
  10.1103/PhysRevA.71.044101} {\bibfield  {journal} {\bibinfo  {journal} {Phys.
  Rev. A}\ }\textbf {\bibinfo {volume} {71}},\ \bibinfo {pages} {044101}
  (\bibinfo {year} {2005})}\BibitemShut {NoStop}%
\bibitem [{\citenamefont {Yi}\ \emph {et~al.}(2006)\citenamefont {Yi},
  \citenamefont {Tong}, \citenamefont {Wang}, \citenamefont {Kwek},\ and\
  \citenamefont {Oh}}]{yi_geometric_2006}%
  \BibitemOpen
  \bibfield  {author} {\bibinfo {author} {\bibfnamefont {X.~X.}\ \bibnamefont
  {Yi}}, \bibinfo {author} {\bibfnamefont {D.~M.}\ \bibnamefont {Tong}},
  \bibinfo {author} {\bibfnamefont {L.~C.}\ \bibnamefont {Wang}}, \bibinfo
  {author} {\bibfnamefont {L.~C.}\ \bibnamefont {Kwek}}, \ and\ \bibinfo
  {author} {\bibfnamefont {C.~H.}\ \bibnamefont {Oh}},\ }\href {\doibase
  10.1103/PhysRevA.73.052103} {\bibfield  {journal} {\bibinfo  {journal} {Phys.
  Rev. A}\ }\textbf {\bibinfo {volume} {73}},\ \bibinfo {pages} {052103}
  (\bibinfo {year} {2006})}\BibitemShut {NoStop}%
\bibitem [{\citenamefont {Tanimura}\ and\ \citenamefont
  {Kubo}(1989)}]{tanimura_1}%
  \BibitemOpen
  \bibfield  {author} {\bibinfo {author} {\bibfnamefont {Y.}~\bibnamefont
  {Tanimura}}\ and\ \bibinfo {author} {\bibfnamefont {R.}~\bibnamefont
  {Kubo}},\ }\href {\doibase 10.1143/JPSJ.58.101} {\bibfield  {journal}
  {\bibinfo  {journal} {J. Phys. Soc. Jpn.}\ }\textbf {\bibinfo {volume}
  {58}},\ \bibinfo {pages} {101} (\bibinfo {year} {1989})}\BibitemShut
  {NoStop}%
\bibitem [{\citenamefont {Tanimura}(1990)}]{tanimura_2}%
  \BibitemOpen
  \bibfield  {author} {\bibinfo {author} {\bibfnamefont {Y.}~\bibnamefont
  {Tanimura}},\ }\href {\doibase 10.1103/PhysRevA.41.6676} {\bibfield
  {journal} {\bibinfo  {journal} {Phys. Rev. A}\ }\textbf {\bibinfo {volume}
  {41}},\ \bibinfo {pages} {6676} (\bibinfo {year} {1990})}\BibitemShut
  {NoStop}%
\bibitem [{\citenamefont {Tanaka}\ and\ \citenamefont
  {Tanimura}(2009)}]{tanimura_3}%
  \BibitemOpen
  \bibfield  {author} {\bibinfo {author} {\bibfnamefont {M.}~\bibnamefont
  {Tanaka}}\ and\ \bibinfo {author} {\bibfnamefont {Y.}~\bibnamefont
  {Tanimura}},\ }\href {\doibase 10.1143/JPSJ.78.073802} {\bibfield  {journal}
  {\bibinfo  {journal} {J. Phys. Soc. Jpn.}\ }\textbf {\bibinfo {volume}
  {78}},\ \bibinfo {pages} {073802} (\bibinfo {year} {2009})}\BibitemShut
  {NoStop}%
\bibitem [{\citenamefont {Tanimura}\ and\ \citenamefont
  {Mukamel}(1994)}]{tanimura_4}%
  \BibitemOpen
  \bibfield  {author} {\bibinfo {author} {\bibfnamefont {Y.}~\bibnamefont
  {Tanimura}}\ and\ \bibinfo {author} {\bibfnamefont {S.}~\bibnamefont
  {Mukamel}},\ }\href {\doibase 10.1143/JPSJ.63.66} {\bibfield  {journal}
  {\bibinfo  {journal} {J. Phys. Soc. Jpn.}\ }\textbf {\bibinfo {volume}
  {63}},\ \bibinfo {pages} {66} (\bibinfo {year} {1994})}\BibitemShut {NoStop}%
\bibitem [{\citenamefont {Tanimura}\ and\ \citenamefont
  {Wolynes}(1991)}]{tanimura_5}%
  \BibitemOpen
  \bibfield  {author} {\bibinfo {author} {\bibfnamefont {Y.}~\bibnamefont
  {Tanimura}}\ and\ \bibinfo {author} {\bibfnamefont {P.~G.}\ \bibnamefont
  {Wolynes}},\ }\href {\doibase 10.1103/PhysRevA.43.4131} {\bibfield  {journal}
  {\bibinfo  {journal} {Phys. Rev. A}\ }\textbf {\bibinfo {volume} {43}},\
  \bibinfo {pages} {4131} (\bibinfo {year} {1991})}\BibitemShut {NoStop}%
\bibitem [{\citenamefont {Tanaka}\ and\ \citenamefont
  {Tanimura}(2010)}]{tanimura_6}%
  \BibitemOpen
  \bibfield  {author} {\bibinfo {author} {\bibfnamefont {M.}~\bibnamefont
  {Tanaka}}\ and\ \bibinfo {author} {\bibfnamefont {Y.}~\bibnamefont
  {Tanimura}},\ }\href {\doibase doi:10.1063/1.3428674} {\ \textbf {\bibinfo
  {volume} {132}},\ \bibinfo {pages} {214502} (\bibinfo {year}
  {2010})}\BibitemShut {NoStop}%
\bibitem [{\citenamefont {Ishizaki}\ and\ \citenamefont
  {Tanimura}(2005)}]{tanimura_7}%
  \BibitemOpen
  \bibfield  {author} {\bibinfo {author} {\bibfnamefont {A.}~\bibnamefont
  {Ishizaki}}\ and\ \bibinfo {author} {\bibfnamefont {Y.}~\bibnamefont
  {Tanimura}},\ }\href {\doibase 10.1143/JPSJ.74.3131} {\bibfield  {journal}
  {\bibinfo  {journal} {J. Phys. Soc. Jpn.}\ }\textbf {\bibinfo {volume}
  {74}},\ \bibinfo {pages} {3131} (\bibinfo {year} {2005})}\BibitemShut
  {NoStop}%
\bibitem [{\citenamefont {Ishizaki}\ and\ \citenamefont
  {Tanimura}(2006)}]{tanimura_8}%
  \BibitemOpen
  \bibfield  {author} {\bibinfo {author} {\bibfnamefont {A.}~\bibnamefont
  {Ishizaki}}\ and\ \bibinfo {author} {\bibfnamefont {Y.}~\bibnamefont
  {Tanimura}},\ }\href {\doibase doi:10.1063/1.2244558} {\ \textbf {\bibinfo
  {volume} {125}},\ \bibinfo {pages} {084501} (\bibinfo {year}
  {2006})}\BibitemShut {NoStop}%
\bibitem [{\citenamefont {Ishizaki}\ and\ \citenamefont
  {Tanimura}(2007)}]{tanimura_9}%
  \BibitemOpen
  \bibfield  {author} {\bibinfo {author} {\bibfnamefont {A.}~\bibnamefont
  {Ishizaki}}\ and\ \bibinfo {author} {\bibfnamefont {Y.}~\bibnamefont
  {Tanimura}},\ }\href {\doibase 10.1021/jp072880a} {\bibfield  {journal}
  {\bibinfo  {journal} {J. Phys. Chem. A}\ }\textbf {\bibinfo {volume} {111}},\
  \bibinfo {pages} {9269} (\bibinfo {year} {2007})}\BibitemShut {NoStop}%
\bibitem [{\citenamefont {Tanimura}(2006)}]{tanimura_stochastic_2006}%
  \BibitemOpen
  \bibfield  {author} {\bibinfo {author} {\bibfnamefont {Y.}~\bibnamefont
  {Tanimura}},\ }\href {\doibase 10.1143/JPSJ.75.082001} {\bibfield  {journal}
  {\bibinfo  {journal} {J. Phys. Soc. Jpn.}\ }\textbf {\bibinfo {volume}
  {75}},\ \bibinfo {pages} {082001} (\bibinfo {year} {2006})}\BibitemShut
  {NoStop}%
\bibitem [{\citenamefont {Ma}\ \emph {et~al.}(2012)\citenamefont {Ma},
  \citenamefont {Sun}, \citenamefont {Wang},\ and\ \citenamefont
  {Nori}}]{ma_entanglement_2012}%
  \BibitemOpen
  \bibfield  {author} {\bibinfo {author} {\bibfnamefont {J.}~\bibnamefont
  {Ma}}, \bibinfo {author} {\bibfnamefont {Z.}~\bibnamefont {Sun}}, \bibinfo
  {author} {\bibfnamefont {X.}~\bibnamefont {Wang}}, \ and\ \bibinfo {author}
  {\bibfnamefont {F.}~\bibnamefont {Nori}},\ }\href {\doibase
  10.1103/PhysRevA.85.062323} {\bibfield  {journal} {\bibinfo  {journal} {Phys.
  Rev. A}\ }\textbf {\bibinfo {volume} {85}},\ \bibinfo {pages} {062323}
  (\bibinfo {year} {2012})}\BibitemShut {NoStop}%
\bibitem [{\citenamefont {Zhong}\ \emph {et~al.}(2013)\citenamefont {Zhong},
  \citenamefont {Sun}, \citenamefont {Ma}, \citenamefont {Wang},\ and\
  \citenamefont {Nori}}]{zhong_fisher_2013}%
  \BibitemOpen
  \bibfield  {author} {\bibinfo {author} {\bibfnamefont {W.}~\bibnamefont
  {Zhong}}, \bibinfo {author} {\bibfnamefont {Z.}~\bibnamefont {Sun}}, \bibinfo
  {author} {\bibfnamefont {J.}~\bibnamefont {Ma}}, \bibinfo {author}
  {\bibfnamefont {X.}~\bibnamefont {Wang}}, \ and\ \bibinfo {author}
  {\bibfnamefont {F.}~\bibnamefont {Nori}},\ }\href {\doibase
  10.1103/PhysRevA.87.022337} {\bibfield  {journal} {\bibinfo  {journal} {Phys.
  Rev. A}\ }\textbf {\bibinfo {volume} {87}},\ \bibinfo {pages} {022337}
  (\bibinfo {year} {2013})}\BibitemShut {NoStop}%
\bibitem [{\citenamefont {Chen}\ \emph {et~al.}(2010)\citenamefont {Chen},
  \citenamefont {An}, \citenamefont {Tong}, \citenamefont {Luo},\ and\
  \citenamefont {Oh}}]{chen_non-markovian_2010}%
  \BibitemOpen
  \bibfield  {author} {\bibinfo {author} {\bibfnamefont {J.-J.}\ \bibnamefont
  {Chen}}, \bibinfo {author} {\bibfnamefont {J.-H.}\ \bibnamefont {An}},
  \bibinfo {author} {\bibfnamefont {Q.-J.}\ \bibnamefont {Tong}}, \bibinfo
  {author} {\bibfnamefont {H.-G.}\ \bibnamefont {Luo}}, \ and\ \bibinfo
  {author} {\bibfnamefont {C.~H.}\ \bibnamefont {Oh}},\ }\href {\doibase
  10.1103/PhysRevA.81.022120} {\bibfield  {journal} {\bibinfo  {journal} {Phys.
  Rev. A}\ }\textbf {\bibinfo {volume} {81}},\ \bibinfo {pages} {022120}
  (\bibinfo {year} {2010})}\BibitemShut {NoStop}%
\bibitem [{\citenamefont {Bhandari}(2002)}]{bhandari_singularities_2002}%
  \BibitemOpen
  \bibfield  {author} {\bibinfo {author} {\bibfnamefont {R.}~\bibnamefont
  {Bhandari}},\ }\href {\doibase 10.1103/PhysRevLett.89.268901} {\bibfield
  {journal} {\bibinfo  {journal} {Phys. Rev. Lett.}\ }\textbf {\bibinfo
  {volume} {89}},\ \bibinfo {pages} {268901} (\bibinfo {year}
  {2002})}\BibitemShut {NoStop}%
\bibitem [{\citenamefont {Mukunda}\ \emph {et~al.}(2001)\citenamefont
  {Mukunda}, \citenamefont {Arvind}, \citenamefont {Chaturvedi},\ and\
  \citenamefont {Simon}}]{mukunda_bargmann_2001}%
  \BibitemOpen
  \bibfield  {author} {\bibinfo {author} {\bibfnamefont {N.}~\bibnamefont
  {Mukunda}}, \bibinfo {author} {\bibnamefont {Arvind}}, \bibinfo {author}
  {\bibfnamefont {S.}~\bibnamefont {Chaturvedi}}, \ and\ \bibinfo {author}
  {\bibfnamefont {R.}~\bibnamefont {Simon}},\ }\href {\doibase
  10.1103/PhysRevA.65.012102} {\bibfield  {journal} {\bibinfo  {journal} {Phys.
  Rev. A}\ }\textbf {\bibinfo {volume} {65}},\ \bibinfo {pages} {012102}
  (\bibinfo {year} {2001})}\BibitemShut {NoStop}%
\bibitem [{\citenamefont {Simon}\ and\ \citenamefont
  {Mukunda}(1993)}]{simon_bargmann_1993}%
  \BibitemOpen
  \bibfield  {author} {\bibinfo {author} {\bibfnamefont {R.}~\bibnamefont
  {Simon}}\ and\ \bibinfo {author} {\bibfnamefont {N.}~\bibnamefont
  {Mukunda}},\ }\href {\doibase 10.1103/PhysRevLett.70.880} {\bibfield
  {journal} {\bibinfo  {journal} {Physical Review Letters}\ }\textbf {\bibinfo
  {volume} {70}},\ \bibinfo {pages} {880} (\bibinfo {year} {1993})}\BibitemShut
  {NoStop}%
\bibitem [{\citenamefont {Mukunda}\ and\ \citenamefont
  {Simon}(1993)}]{phase_tracking}%
  \BibitemOpen
  \bibfield  {author} {\bibinfo {author} {\bibfnamefont {N.}~\bibnamefont
  {Mukunda}}\ and\ \bibinfo {author} {\bibfnamefont {R.}~\bibnamefont
  {Simon}},\ }\href {\doibase http://dx.doi.org/10.1006/aphy.1993.1093}
  {\bibfield  {journal} {\bibinfo  {journal} {Annals of Physics}\ }\textbf
  {\bibinfo {volume} {228}},\ \bibinfo {pages} {205 } (\bibinfo {year}
  {1993})}\BibitemShut {NoStop}%
\bibitem [{\citenamefont {Scully}\ and\ \citenamefont
  {Zubairy}(1997)}]{quantum_optics_1997}%
  \BibitemOpen
  \bibfield  {author} {\bibinfo {author} {\bibfnamefont {M.~O.}\ \bibnamefont
  {Scully}}\ and\ \bibinfo {author} {\bibfnamefont {M.~S.}\ \bibnamefont
  {Zubairy}},\ }\href@noop {} {\emph {\bibinfo {title} {Quantum Optics}}}\
  (\bibinfo  {publisher} {Cambridge University Press},\ \bibinfo {address}
  {Cambridge},\ \bibinfo {year} {1997})\BibitemShut {NoStop}%
\bibitem [{\citenamefont {Breuer}\ and\ \citenamefont
  {Petruccione}(2002)}]{breuer_open_system}%
  \BibitemOpen
  \bibfield  {author} {\bibinfo {author} {\bibfnamefont {H.-P.}\ \bibnamefont
  {Breuer}}\ and\ \bibinfo {author} {\bibfnamefont {F.}~\bibnamefont
  {Petruccione}},\ }\href@noop {} {\emph {\bibinfo {title} {The Theory of Open
  Quantum Systems}}}\ (\bibinfo  {publisher} {Oxford University Press},\
  \bibinfo {address} {Oxford},\ \bibinfo {year} {2002})\BibitemShut {NoStop}%
\bibitem [{\citenamefont {Intravaia}\ \emph {et~al.}(2003)\citenamefont
  {Intravaia}, \citenamefont {Maniscalco},\ and\ \citenamefont
  {Messina}}]{intravaia_comparison_2003}%
  \BibitemOpen
  \bibfield  {author} {\bibinfo {author} {\bibfnamefont {F.}~\bibnamefont
  {Intravaia}}, \bibinfo {author} {\bibfnamefont {S.}~\bibnamefont
  {Maniscalco}}, \ and\ \bibinfo {author} {\bibfnamefont {A.}~\bibnamefont
  {Messina}},\ }\href {\doibase 10.1140/epjb/e2003-00078-6} {\bibfield
  {journal} {\bibinfo  {journal} {The European Physical Journal B - Condensed
  Matter and Complex Systems}\ }\textbf {\bibinfo {volume} {32}},\ \bibinfo
  {pages} {97} (\bibinfo {year} {2003})}\BibitemShut {NoStop}%
\bibitem [{\citenamefont {M{\"a}kel{\"a}}\ and\ \citenamefont
  {M{\"o}tt{\"o}nen}(2013)}]{makela_effects_2013}%
  \BibitemOpen
  \bibfield  {author} {\bibinfo {author} {\bibfnamefont {H.}~\bibnamefont
  {M{\"a}kel{\"a}}}\ and\ \bibinfo {author} {\bibfnamefont {M.}~\bibnamefont
  {M{\"o}tt{\"o}nen}},\ }\href {\doibase 10.1103/PhysRevA.88.052111} {\bibfield
   {journal} {\bibinfo  {journal} {Physical Review A}\ }\textbf {\bibinfo
  {volume} {88}},\ \bibinfo {pages} {052111} (\bibinfo {year}
  {2013})}\BibitemShut {NoStop}%
\bibitem [{\citenamefont {Leggett}\ \emph {et~al.}(1987)\citenamefont
  {Leggett}, \citenamefont {Chakravarty}, \citenamefont {Dorsey}, \citenamefont
  {Fisher}, \citenamefont {Garg},\ and\ \citenamefont
  {Zwerger}}]{leggett_dynamics_1987}%
  \BibitemOpen
  \bibfield  {author} {\bibinfo {author} {\bibfnamefont {A.~J.}\ \bibnamefont
  {Leggett}}, \bibinfo {author} {\bibfnamefont {S.}~\bibnamefont
  {Chakravarty}}, \bibinfo {author} {\bibfnamefont {A.~T.}\ \bibnamefont
  {Dorsey}}, \bibinfo {author} {\bibfnamefont {M.~P.~A.}\ \bibnamefont
  {Fisher}}, \bibinfo {author} {\bibfnamefont {A.}~\bibnamefont {Garg}}, \ and\
  \bibinfo {author} {\bibfnamefont {W.}~\bibnamefont {Zwerger}},\ }\href
  {\doibase 10.1103/RevModPhys.59.1} {\bibfield  {journal} {\bibinfo  {journal}
  {Reviews of Modern Physics}\ }\textbf {\bibinfo {volume} {59}},\ \bibinfo
  {pages} {1} (\bibinfo {year} {1987})}\BibitemShut {NoStop}%
\bibitem [{\citenamefont {Chin}\ \emph {et~al.}(2011)\citenamefont {Chin},
  \citenamefont {Prior}, \citenamefont {Huelga},\ and\ \citenamefont
  {Plenio}}]{chin_generalized_2011}%
  \BibitemOpen
  \bibfield  {author} {\bibinfo {author} {\bibfnamefont {A.~W.}\ \bibnamefont
  {Chin}}, \bibinfo {author} {\bibfnamefont {J.}~\bibnamefont {Prior}},
  \bibinfo {author} {\bibfnamefont {S.~F.}\ \bibnamefont {Huelga}}, \ and\
  \bibinfo {author} {\bibfnamefont {M.~B.}\ \bibnamefont {Plenio}},\ }\href
  {\doibase 10.1103/PhysRevLett.107.160601} {\bibfield  {journal} {\bibinfo
  {journal} {Physical Review Letters}\ }\textbf {\bibinfo {volume} {107}},\
  \bibinfo {pages} {160601} (\bibinfo {year} {2011})}\BibitemShut {NoStop}%
\bibitem [{\citenamefont {ZHU}(2008)}]{zhu_geometric_2008}%
  \BibitemOpen
  \bibfield  {author} {\bibinfo {author} {\bibfnamefont {S.-L.}\ \bibnamefont
  {ZHU}},\ }\href {\doibase 10.1142/S0217979208038855} {\bibfield  {journal}
  {\bibinfo  {journal} {International Journal of Modern Physics B}\ }\textbf
  {\bibinfo {volume} {22}},\ \bibinfo {pages} {561} (\bibinfo {year}
  {2008})}\BibitemShut {NoStop}%
\end{thebibliography}%
\end{document}